\documentclass[prb,twocolumn,showpacs]{revtex4-1}
	\usepackage{graphicx,epsf,bm,bbm,color,amsmath}
\usepackage{ulem}

	\newcommand{\maH}{\mathcal{H}}
	\newcommand{\beq}{\begin{equation}}
	\newcommand{\beqn}{\begin{eqnarray}}
	\newcommand{\eeq}{\end{equation}}
	\newcommand{\eeqn}{\end{eqnarray}}
	\newcommand{\nn}{\nonumber}
	\newcommand{\bq}{{\bf q}}
	\newcommand{\be}{{\bf e}}
	\newcommand{\ba}{{\bf a}}
	\newcommand{\bk}{{\bf k}}
	\newcommand{\bK}{{\bf K}}
	
	\newcommand{\ep}{{\epsilon}}

\begin{document}

		\title{Magnetic spectrum of trigonally warped bilayer graphene -- semiclassical analysis, zero modes, and topological winding numbers}
		\author{R. de Gail, M. O. Goerbig and G. Montambaux}
		\affiliation{Laboratoire de Physique des Solides, CNRS UMR 8502, Univ. Paris-Sud, F-91405 Orsay cedex, France.}

		\begin{abstract}
We investigate the fine structure in the energy spectrum of bilayer graphene in the presence of various stacking defaults, such as
a translational or rotational mismatch. This fine structure consists of four Dirac points that move away from their
original positions as a consequence of the mismatch and eventually merge in various manners. The different types of merging are
described in terms of topological invariants (winding numbers) that determine the Landau-level spectrum in the presence of a magnetic
field as well as the degeneracy of the levels. The Landau-level spectrum is, within a wide parameter range, well described by a
semiclassical treatment that makes use of topological winding numbers. However, the latter need to be redefined at zero energy in the
high-magnetic-field limit as well as in the vicinity of saddle points in the zero-field dispersion relation. 

		\end{abstract}
	\maketitle
	
	\section{Introduction}

Graphene research has stimulated many fields of condensed-matter physics during the last years.\cite{AntonioReview} One of the most
remarkable of these fields is certainly the topological description of electronic energy bands, such as in the context of topological
insulators.\cite{HasanRev10,ZhangRev} Indeed, the low-energy electronic properties of a single graphene layer
are determined by two particular band-contact
points at the corners $K$ and $K'$ of the first Brillouin zone, with a  linear dispersion relation (the so-called Dirac
points). These Dirac points are associated with a topological Berry phase that stems from the winding of the phase in the electronic
wave function on closed paths around these points -- the Berry phase is then $\pi$ times this winding number.\cite{Fuchs10} Prominent
consequences of this Berry phase are the absence of backscattering in the case of long-range disorder\cite{AntonioReview}, Klein
tunneling,\cite{katsnelson06,AllainFuchs} and a particular form of the Landau level (LL) spectrum in the presence of a magnetic field,
with a topologically protected zero-energy level.\cite{GoerbigReview}

Bilayer graphene, that is obtained from an AB stacking of two graphene layers, has an even richer band structure, also from a topological
point of view, than monolayer graphene. Most of its electronic properties have successfully been described in the framework
of two parabolic bands with opposite curvature that touch each other at the Fermi level. As compared to monolayer graphene, the winding
number associated with these band-contact points is twice as large.\cite{MF06} This gives rise to a two-fold orbital degeneracy of the
zero-energy level in the presence of a magnetic field, in addition to the four-fold spin-valley degeneracy, and thus to a particular
series of Hall plateaus that have been observed in quantum-Hall measurements.\cite{bilayerQHE} However, this picture is only approximately
valid in an intermediate energy range (above $\simeq 10$ meV), whereas subordinate hopping terms yield a fine structure in the energy
spectrum (called ``trigonal warping''), in the form of four Dirac points with linear dispersion, at lower energies.\cite{MF06} This
transition from four Dirac points, with unit winding numbers, to the parabolic regime with a winding number of 2 may be viewed as
a finite-energy Lifshitz transition\cite{Lifshitz} between disconnected Fermi pockets at low energies and a simply connected Fermi sea
(per valley) at higher energies.\cite{Lemonik}
In contrast to earlier experiments on bilayer graphene, today's availability of high-quality samples allows one to probe
now this low-energy regime in which quantum-Hall measurements indicate the presence of additional Dirac points,\cite{Novoselov11}
and it is noteworth to mention that indications of Lifshitz transitions have been found in cyclotron-resonance measurements in 
graphite.\cite{orlita}

The fine structure of the energy spectrum of bilayer graphene is also interesting from the point of view of stacking defaults, such as
a displacement, strain or a twist with respect to perfect AB stacking.\cite{Lopes,Son,MacDonald10a} In this case, the low-energy
dispersion is modified and two or more of the Dirac points may easily merge.\cite{Son,degail11} This needs to be contrasted to
Dirac-point merging in monolayer graphene that has been extensively studied on the theoretical level\cite{hasegawa,zhu,dietl,wunsch,Gilles2,Gilles1}
but that is difficult
to achieve experimentally due to an enormous strain required.\cite{strain} Furthermore, moderate stacking defaults may allow for the
systematic study of merging transitions that fall into two distinct topological classes\cite{degail11} -- whereas the merging of
Dirac points with opposite winding numbers yields a gap in the band structure, that of Dirac points with the same winding number
maintains the band-contact points. This difference has direct consequences for the LL spectrum, namely the zero-energy level.
Whereas in the former case of merging Dirac points with opposite winding number, the twofold degeneracy of the zero-energy level
is lifted,\cite{dietl} it is topologically protected in the latter case.\cite{degail11}

Here, we investigate the different merging transitions that one may encounter in bilayer graphene with a stacking default, within
a continuum model that has been used both in the description of bilayer graphene with a mismatch described by a translation between the
layers or under strain\cite{Son,Falko11} as well as in that of a twisted bilayer.\cite{degail11} This continuum model, which goes beyond
the linear Dirac-point approximation, may be viewed as a continuum model of the second generation.\cite{MeleReview} In addition to the
merging transition between Dirac points of opposite winding number, we discuss in detail the triple merging of three Dirac points
that has been investigated in previous theoretical works.\cite{Son,Falko11,ecrys} This triple merging happens to be unstable in the sense that it
only occurs in the framework of a displacement or strain in a high-symmetry axis of the lattice -- a slight deviation from such an axis
splits the triple-merging into a usual merging transition of two Dirac points in a first step, followed by merging with the remaining
Dirac point in a second step. 
As compared to previous studies of the LL spectrum for trigonally warped bilayer graphene,\cite{MF06,Falko11} we provide in the present paper a
detailed semiclassical analysis of the spectrum. This analysis is based on winding numbers that allow for a transparent understanding of
LL degeneracies and zero modes. Furthermore, we investigate quantum corrections beyond the semiclassical limit.
Both at zero energy in the high-field limit and in the vicinity of saddle points in the dispersion relation, these
corrections are relevant because they blur the semiclassical trajectories
and thus call for a modification of the description in terms of winding numbers. This allows for an understanding
of the change in the LL degeneracy at zero energy and in the vicinity of saddle points in the dispersion relation.

  The paper is organized as follows. In Sec. \ref{sec:bands}, we present the continuum model that accounts for the different
stacking defaults in bilayer graphene and discuss the band structure and zero-field merging transitions in the several limits.
Furthermore, we characterize the merging transitions in terms of topological winding numbers. Section
\ref{sec:LL} is devoted to the LL spectrum associated with the different merging transitions. The spectrum is obtained within a numerical
solution of the quantum-mechanical eigenvalue equation (Sec. \ref{sec:LQ}) and analyzed in the framework of a semiclassical treatment
(Sec. \ref{sec:semicl}). Topological aspects of the LL spectrum are discussed in Sec. \ref{sec:topology}, and a detailed discussion of
the different aspects of the spectrum may be found in Sec. \ref{sec:discussion}, before we present our conclusions
(Sec. \ref{sec:concl}).

	\section{Band Structure of Deformed Bilayer Graphene}\label{sec:bands}
\begin{figure}[h!]
	\includegraphics[width=1.0 \columnwidth]{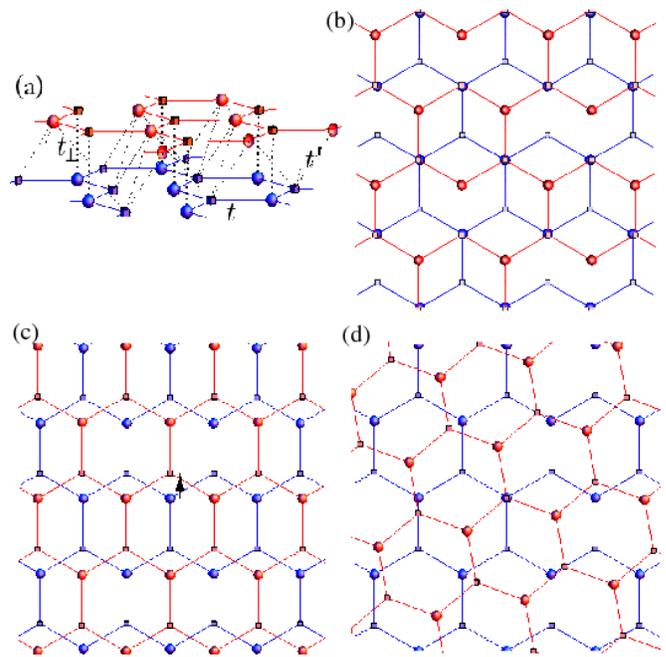}
	\caption{(Color online) Lattice Structure of the Bilayer Graphene.
	(a) Atomic structure around an elementary cell for the Bernal configuration with ($A$,$B$) atoms  (first layer) in blue and  ($\tilde{A},\tilde{B}$) atoms (second layer) in red.  Hopping parameters $t$, $t_\perp$ and $t'$ are also pictured.
	Figures (b), (c) and (d) depict a planar lattice geometry for the Bernal, slided and twisted bilayer, respectively. }
	\label{lattice}
	\end{figure}

Bilayer graphene harbors different stacking geometries, pictured in Fig. \ref{lattice},
among which the energetically most favorable is the Bernal (AB) configuration in which
a $B$ sublattice atom of the first layer sits on top of an $\tilde{A}$ sublattice atom of the second layer [Fig. \ref{lattice}(a)].
This particular ordering is naturally observed in graphite, as well as for synthesized bilayer graphene.
Other stackings  may be described in terms of a rotation default\cite{MeleReview} and a displacement vector\cite{Son,Falko11} and may be observed in graphene samples, such as for example in epitaxial graphene on the C-face of the SiC crystal.\cite{Hass08}
We also consider strain constraints along both layers.\cite{Falko11,Falko12}

			\subsection{Tight-Binding Approach}
For perfect AB-stacking,\cite{MF06} the tight-binding approximation yields a
four-band Hamiltonian that may be written in the $(A,B,\tilde{A},\tilde{B})$ basis
	\beq
	\label{Hamlattice}
	\maH(\bk) =
	\left( \begin{array}{cccc}
	0 & t\gamma(\bk) & 0 & t' \gamma^*(\bk) \\
	t\gamma^*(\bk) & 0 & t_{\perp} & 0 \\
	0 & t_{\perp} & 0 & t\gamma(\bk) \\
	t' \gamma(\bk) & 0 & t\gamma^*(\bk) & 0
	\end{array} \right),
	\eeq
where
	\beq
	\gamma (\bk)  =  -(1 +  e^{i\bk \cdot \ba_1}+e^{i\bk \cdot \ba_2} ),
	\eeq
and $\ba_1$, $\ba_2$ are elementary vectors of the triangular Bravais lattice
	\beq
	\ba_{1,2}  =  \frac{a}{2}(\pm\sqrt{3}\be_x+ 3 \be_y)
	\eeq
and $a = 0.142$ nm is the distance between neighboring carbon atoms in the same layer.
The hopping parameters can be experimentally evaluated,\cite{Zou11} and one obtains the hierarchy
	\beq\label{hoppingOM}
	t \, (\sim 3 \, \text{eV}) \gg t_{\perp} \,(\sim 0.4 \, \text{eV}) \gtrsim t' \, (\sim \, 0.3 \, \text{eV}),
	\eeq
where $t$ represents the hopping between $p_z$ orbitals of nearest-neighbor carbon atoms within the same layer, $t_{\perp}$ the perpendicular hopping 
amplitude between a $B$ sublattice atom of one layer and the $\tilde{A}$ atom of the other layer, and $t'$ is the transfer integral from an $A$ site 
of one layer to the nearest $\tilde{B}$ sites of the other layer  [see Fig. \ref{lattice}(a)].
All other orbital overlap may be neglected for energies larger than $1$ meV.\cite{Zou11}
		
		\subsection{Low-Energy Hamiltonian}
In the small-wave-vector limit ($|\bq|\ll 1/a$), Hamiltonian (\ref{Hamlattice}) may be expanded around a $K$ or $K'$ corner of the hexagonal
Brillouin zone situated at the positions $\pm \bK=\pm 4\pi\be_x/3\sqrt{3}a$, modulo a reciprocal lattice vector.
One has then $t\gamma(\pm\bK+\bq) \approx v_F (\pm q_x- iq_y)$, in terms of the Fermi
velocity $v_F=3ta/2$  ($\hbar = 1$ henceforth) and $q \ll K$.
Furthermore, for energies lower than $t_{\perp}$, only two bands are relevant, and they may be described with the help of an
effective two-band continuum Hamiltonian\cite{MF06}
	\beq
	\label{HamK}
	\maH_K \approx
	b \left( \begin{array}{cc} 0 &   \pi^{\dagger2}  \\   \pi^{2} & 0 \end{array} \right)
	+
	c \left( \begin{array}{cc} 0 &   \pi \\   \pi^{\dagger}  & 0 \end{array} \right)
	= \maH_b + \maH_c,
	\eeq
Here, $|b|=  v_F^2/t_\perp \approx 14/m_0$, in terms of the  bare electron mass $m_0$,
$c = v_F t'/t \approx 10^5\,\text{m/s}$, and $\pi=q_x+iq_y$ is the complex momentum operator
in the continuum limit (that changes as $\pi \rightarrow - \pi^\dagger$ when $\bK \rightarrow \bK'=-\bK$).
The $\maH_b$ term in Eq. (\ref{HamK}) is dominant for energies higher than $\sim 10$ meV and lower than $t_\perp \sim 0.4$ eV.
In the absence of the term $\maH_c$, it enforces a quadratic dispersion around the band-contact points at $K$ and $K'$.
For energies lower than $\sim 10$ meV, $\maH_c$ becomes relevant and trigonally warps the band structure, which now presents
four Dirac cones  (see Fig. \ref{band0}). One of the Dirac points ($D$) remains at the center $\bq=0$,
whereas three additional cones ($A$, $B$, and $C$) are arranged in a triangle around the first one.

			\subsubsection{Slide and strain deformation}
The translational and strain constraints may be accounted for by adding a constant shift
\beq
\label{Delta}
\maH_{\Delta} =
\left( \begin{array}{cc} 0 & -\Delta \\ -\Delta^* & 0 \end{array} \right),
\eeq
to Hamiltonian (\ref{HamK}) that represents the only relevant perturbation whenever
time-reversal and lattice-inversion symmetries are preserved.\cite{Manes}
Hence translation and strain constraints inevitably give rise to the term (\ref{Delta}).
For instance, a small sliding deformation renders the $t'$ hopping anisotropic due to different orbital overlaps.
In a similar fashion to the anisotropic honeycomb lattice problem,\cite{Gilles1,Gilles2} the renormalized amplitude modifies the continuum approximation by shifting the momentum by a constant value $\Delta$,
\beq
c\pi \rightarrow c\pi - \Delta.
\eeq
The effective Hamiltonian
\beq\label{fullHam}
\maH_T = \maH_b + \maH_c + \maH_\Delta
\eeq
was introduced in Ref. \onlinecite{Son} to take into account a translational mismatch between the two graphene layers. A more microscopic discussion
of the model may be found in Refs. \onlinecite{Falko11,Falko12}.

			\subsubsection{Rotational default}
In the case of a twisted (or rotationally-faulted) bilayer, lattice-inversion symmetry is broken. 
For small and moderate twist angles, the model $\maH_b+\maH_\Delta$ is an approximation that yields the correct shape of the energy 
spectrum and the right topological properties of the original system, such as the degeneracy of the zero-energy Landau level.\cite{degail11} The full 
band structure requires taking into account the commensurability between the rotated layers and the resulting Moir\'e patterns.\cite{MacDonald10a}
Trigonal warping within the twisted bilayer system is likely to be negligible since the orbital mismatch renders all hopping parameters small 
compared to $t$ or $t_{\perp}$, such that $\maH_c=0$. Notice furthermore that also $t_{\perp}$ is significantly lowered by the twist.
For this particular reason, we do not consider $\maH_T$ in Eq. (\ref{fullHam})
as a universal Hamiltonian for bilayer graphene but rather as a model that correctly interpolates between several configurations 
that exist under various experimental conditions.

Notice that interaction effects generate the same distortion $\maH_{\Delta}$, both with\cite{Lemonik} and without\cite{Vafek} trigonal warping.

		\subsection{Band Structure}\label{sec:BandStruc}
The band structure of $\maH_T$ in Eq. (\ref{fullHam}) is plotted in Fig. \ref{band} for various values of $\Delta$.

			\subsubsection{Undistorted case}
\begin{figure}[h!]
	\includegraphics[width=1\columnwidth]{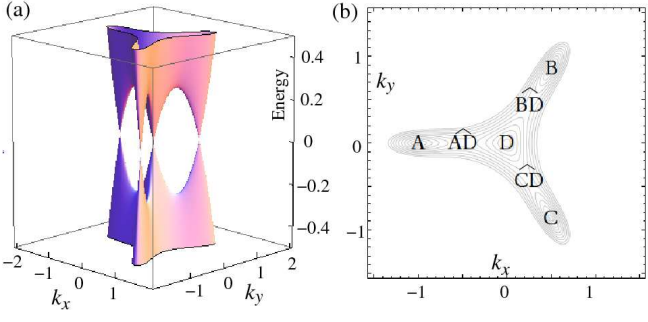}
	\caption{(Color online) (a) Band structure of the perfectly AB-stacked bilayer graphene around one of the valleys. (b) Position of the remarkable points in reciprocal space.
	We label the four Dirac cones from $A$ to $D$ and the corresponding saddle points $\widehat{AD}$, $\widehat{BD}$ and $\widehat{CD}$.
	While the Dirac points all reside at zero energy, the saddle points have an energy $E = c^2/4b$. 
The wave vectors are measured in units of $c/b$ and the energy in units of $c^2/b$.}
	\label{band0}
	\end{figure}

Without any distortion ($\Delta=0$, see Fig. \ref{band0}), the band structure is  trigonally symmetric,
 with a central cone $D$ and three peripheral ones, $A$ $B$ and $C$ positioned at
\beqn\label{eq:DPD0}
\nn
D = (0,0), \quad  A = \left(-\frac{c}{b},0\right) ,\\
B/C = \left(\frac{c}{2b},\pm \frac{\sqrt{3}c}{2b}\right),
\eeqn
within a valley.
A Taylor expansion of the energy dispersion around the four Dirac points yields
\beqn\label{eq:DispD0}
\nn
E_D(\bq) & = &  c \sqrt{q_x^2+q_y^2}, \\
\nn
E_A(\bq) & = & c \sqrt{q_x^2+9 q_y^2}, \\
E_{B/C}(\bq) & = & c \sqrt{7q_x^2+3q_y^2 \pm 4\sqrt{3}q_x q_y},
\eeqn
such that one may define averaged Fermi velocities, that is $v_D=\sqrt{v_x v_y} = c$ for the $D$ cone and $v_A=v_B=v_C=\sqrt{3}c$
for the satellite ones.\cite{MF06}
Three saddle points join each peripheral cone to the central one, see Fig. \ref{band0}, and are located at
\beqn\label{eq:SPD0}
\nn
\widehat{AD} &  = & \left(-\frac{c}{2b},0\right),  \\
\widehat{BD}/\widehat{CD} & = & \left(\frac{c}{4b},\pm \frac{\sqrt{3}c}{4b}\right).
\eeqn
As a consequence of the trigonal symmetry, they occur all at the same energy
\beqn
\nn
E_{\widehat{AD}} & = & E_S = \frac{c^2}{4b}, \\
E_{\widehat{BD}} & = & E_{\widehat{CD}} = E_{S'} = \frac{c^2}{4b},
\label{saddle-point}
\eeqn
that is 
\beq
E_S=E_{S'}=\frac{1}{4}\left(\frac{t'}{t}\right)^2t_{\perp}\simeq 1\, \text{meV},
\eeq
where we have used the values of Eq. (\ref{hoppingOM}) for the hopping amplitudes.

			\subsubsection{Deformation along an axis of high symmetry}
The trigonal point-symmetry is broken as soon as $\Delta \neq 0$.
	\begin{figure}[!h]
	\includegraphics[width=1.0\columnwidth]{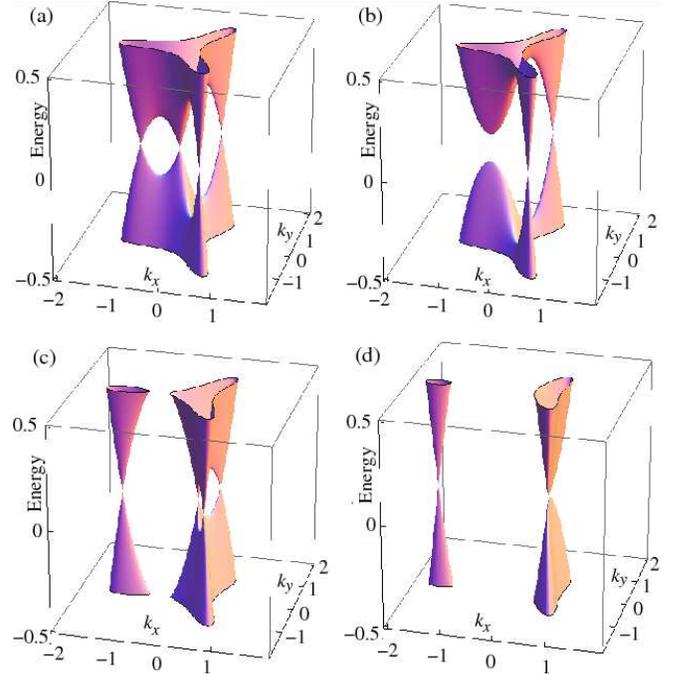}
	\caption{ (Color online) Band structure of bilayer graphene around the $K$ valley for the Hamiltonian $\maH_{\Delta}$, with a real value of 
$\Delta$.
For $\Delta <0$, the two cones $D$ and $A$ start to merge [panels (a), for $\Delta=-0.1c^2/b$], and give rise to a local minimum after the merging
transition, at $\Delta=-c^2/4b$, [panel (b) for a value of $\Delta=-0.32 c^2/b$]. 
The opposite case of $\Delta > 0$ [panels (c) for $\Delta=0.3 c^2/b$ and (d) for $\Delta=0.92 c^2/b$], reveals the merging of three cones at a time, 
$D$ $B$ and $C$, or triple merging. 
The wave vectors are measured in units of $c/b$ and the energy in units of $c^2/b$.}
	\label{band}
	\end{figure}
We first consider the case of a real-valued constant, corresponding to an applied deformation along the $y$-axis [see Fig. \ref{lattice}(c)],
that is an axis of high symmetry.
The Dirac cones are then moved from the positions (\ref{eq:DPD0}) to
\beqn
\nn
D & = & \left(-\frac{c-\sqrt{c^2+4b\Delta}}{2b},0\right),\\
\nn
A &=& \left(-\frac{c+\sqrt{c^2+4b\Delta}}{2b},0\right), \\
B/C & = &\left(\frac{c}{2b},\pm \sqrt{\frac{3c^2}{4b^2}-\frac{\Delta}{b}}\right),
\eeqn
with the averaged Fermi velocities
\beqn\label{eq:FermiV}
\nn
v_D^2 & = & \sqrt{c^2+4b\Delta}\left(2c-\sqrt{c^2+4b\Delta}\right), \\
\nn
v_A^2 & = & \sqrt{c^2+4b\Delta}\left(2c+\sqrt{c^2+4b\Delta}\right), \\
v_{B/C}^2 & = & \sqrt{(5c^2-4b\Delta)^2-16c^2(c^2-b\Delta)}.
\eeqn

Moreover, the positions of the saddle points are shifted from those described in Eq. (\ref{eq:SPD0}) to
\beqn
\widehat{AD} & = & \left(-\frac{c}{2b},0\right)\\
\nn
\widehat{BD}/\widehat{CD} & = & \left(-\frac{3c^2+4b\Delta}{12bc}, \right. \\
\nn
&&\left. \pm \frac{1}{12bc}\sqrt{(3c^2-4b\Delta)(9c^2+4b\Delta)}\right).
\eeqn
The saddle points $\widehat{BD}$ and $\widehat{CD}$ are at the same energy
\beq
E_{S'}  =  \frac{1}{12\sqrt{3}bc}(3c^2-4b\Delta)^{3/2},
\eeq
whereas that between $\widehat{AD}$ is found at
\beq
E_S  =  \frac{c^2}{4b}+\Delta \neq E_{S'}.
\eeq

\begin{figure}[h!]
	\includegraphics[width=.9\columnwidth]{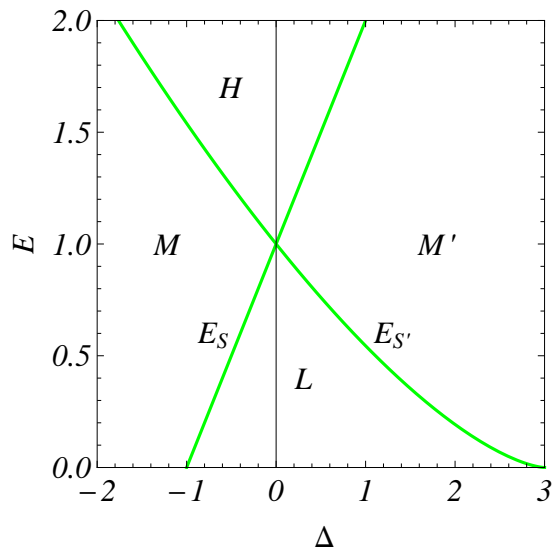}
	\caption{(Color online) In the plane ($E$, $\Delta$), the positions of the two saddle points $E_S$ and $E_{S'}$ define four distinct regions. The
sector L denotes energies that are below both saddle-point energies $E_S$ and $E_{S'}$, whereas H describes energies $E>E_S,E_{S'}$. The sector M is defined
as energies $E$, with $E_S<E<E_{S'}$, and M$^{\prime}$ for $E_{S'}<E<E_S$. Here the energies are given in units of $c^2/4b$. }
	\label{fig:secteurs}
	\end{figure}

Whenever $E_S \neq E_{S'}$, it is possible to define different low-energy regions that make a distinction between the four cones
(Fig. \ref{fig:secteurs}). These regions turn out to be useful for the discussion of the LL spectrum in the Sec. \ref{sec:LL}.
For instance, in the case $\Delta < 0$, $E_S<E_{S'}$ and the $A$ and $D$ cones start to move closer together and eventually merge when $E_S=0$, that is at $\Delta = -c^2/4b$.
The other two cones stay apart, well separated [see Fig. \ref{band}(a)].
 Exactly at the merging transition,
the band dispersion is a semi-Dirac one, quadratic in one direction, linear in the other,   whereas beyond the transition
a gap opens with a quadratic dispersion in both directions [Fig. \ref{band}(b)].
We emphasize that this merging transition between the pair of Dirac cones is exactly the same as in the case  
where the two Dirac cones were related by time-reversal symmetry, as discussed 
in Refs. \onlinecite{Gilles1,Gilles2}.

On the other hand, when $\Delta>0$ one has $E_{S'}<E_{S}$, such that the $D$, $B$ and $C$ cones converge to a common point whereas $A$ stands alone.
The three cones are coupled at energies around $E_{S'}$ [see Fig. \ref{band}(c)].
At the (triple) merging transition, $\Delta = 3c^2/4b$,
the crossing bands bear a complex boomerang shape [Fig. \ref{band}(d)],   while further increase of $\Delta$
does not open a gap in the band structure, in contrast to the above-mentioned merging transition for $\Delta<0$.
This difference may be understood in terms of winding numbers that play the role of topological charges and that are described in
detail in Sec. \ref{topo:zero}.

Notice that, since trigonal warping is a structure at very low energy ($\lesssim 10$ meV),
a small perturbation is sufficient to drive the system into one of the merging scenarios.
For instance, a (triple) merging of the Dirac points occurs at a very small ($\sim 0.10$\r{A}) displacement of one graphene layer with respect to the other one, where we use perfect AB stacking as the reference.\cite{Son}

			\subsubsection{Deformation along an unspecified axis}
\begin{figure}[!h]
	\includegraphics[width=0.45\columnwidth]{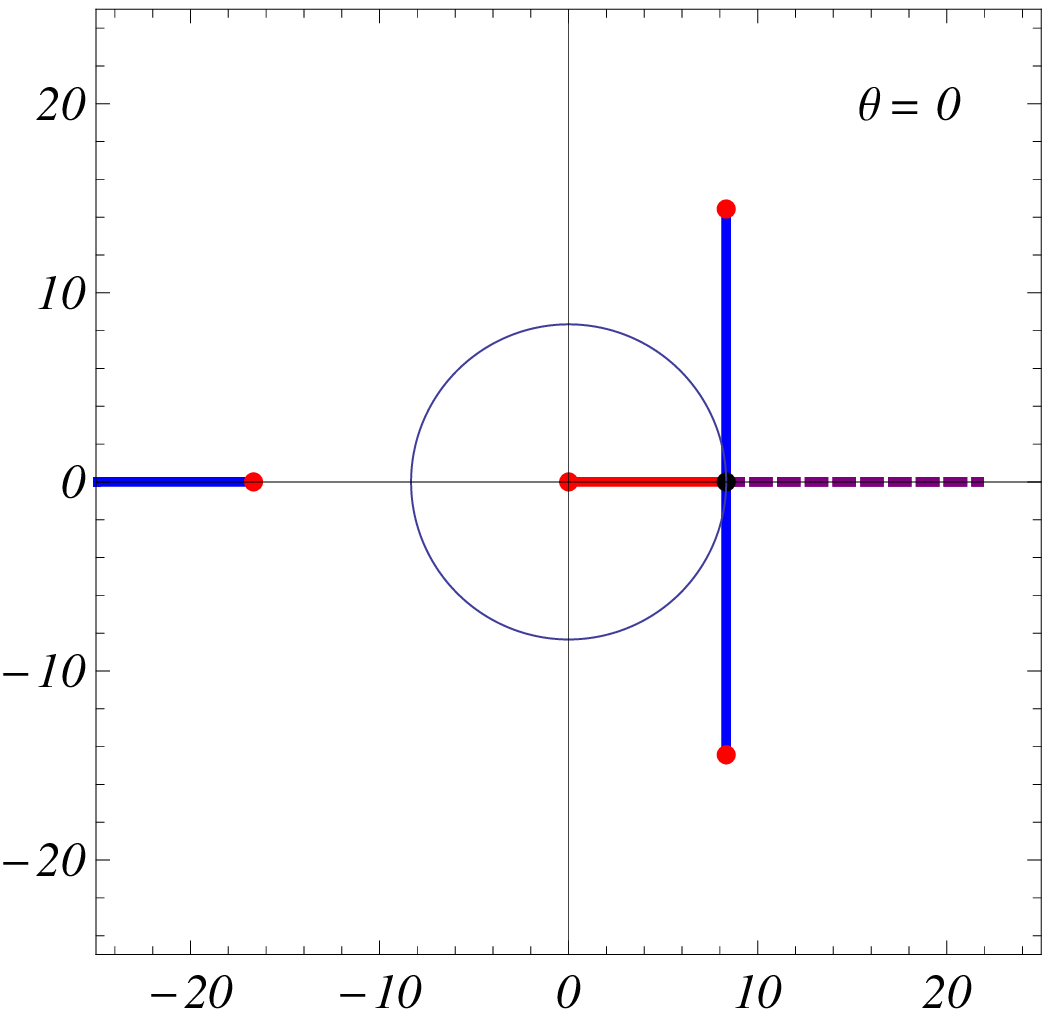}
	\includegraphics[width=0.45\columnwidth]{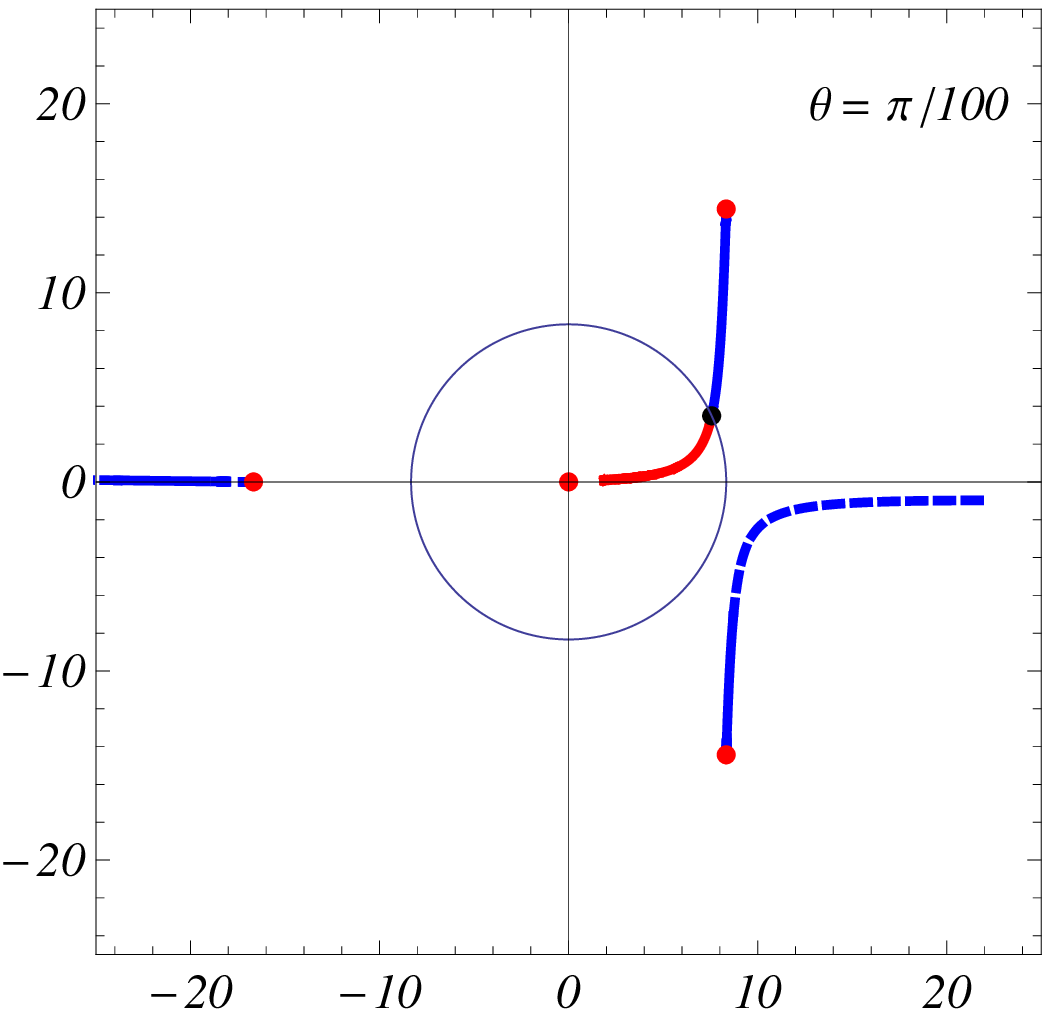}
	\\
	\includegraphics[width=0.45\columnwidth]{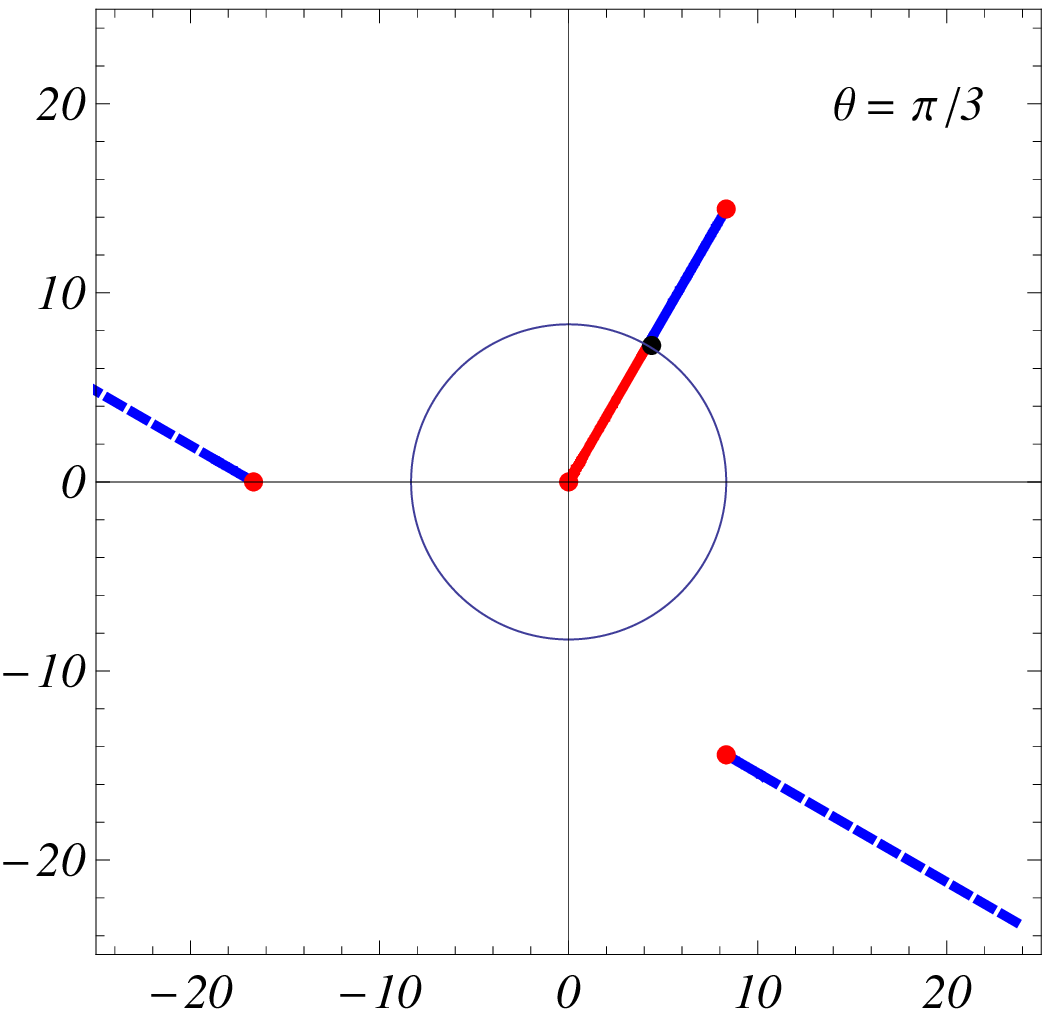}
	\includegraphics[width=0.45\columnwidth]{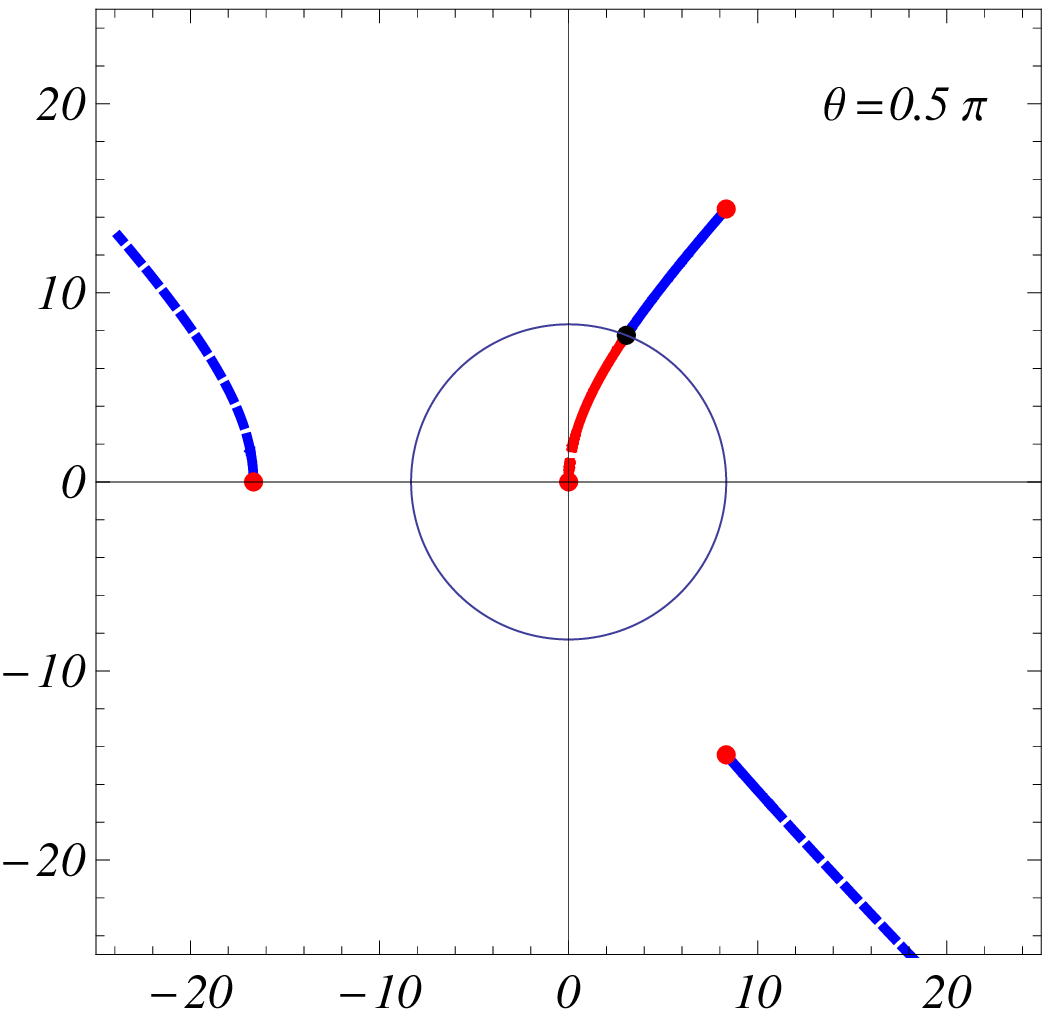}
	\\
	\includegraphics[width=0.45\columnwidth]{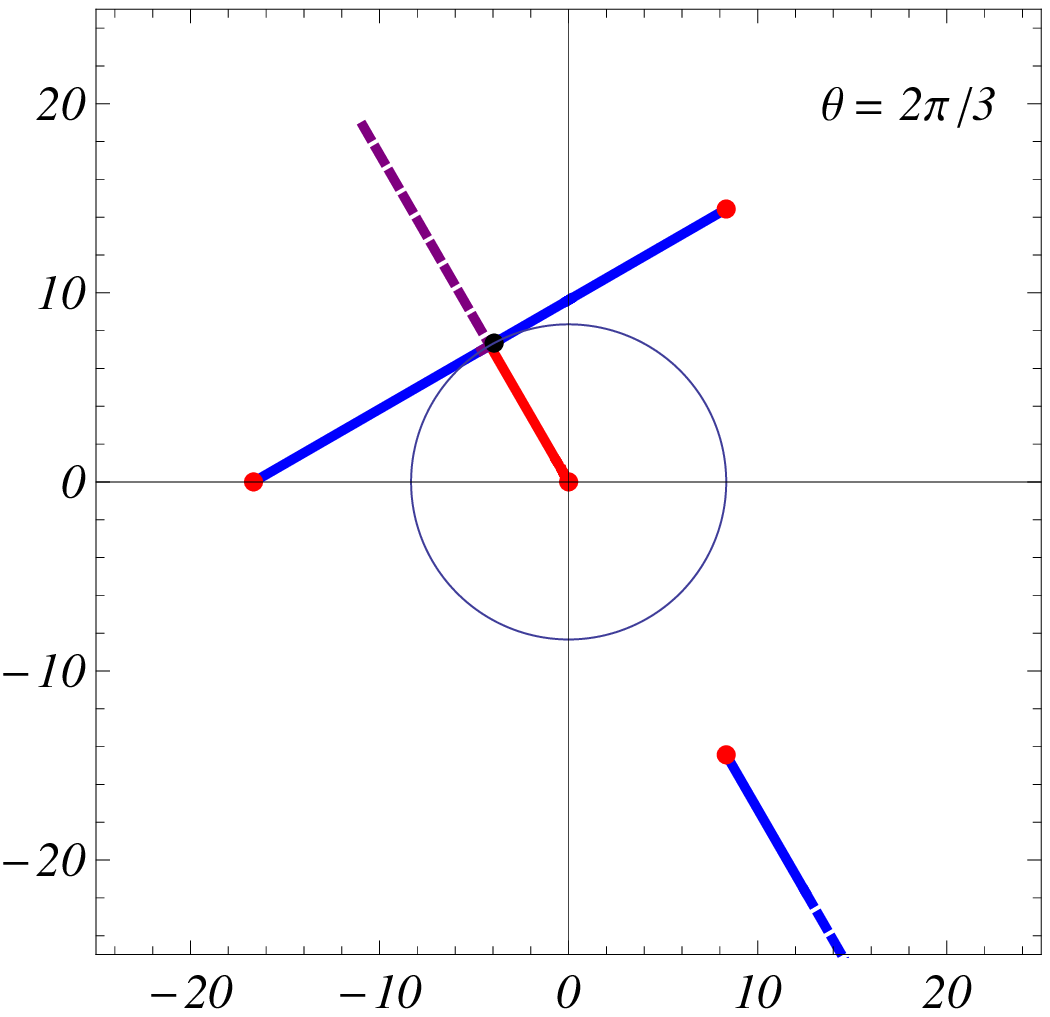}
	\includegraphics[width=0.45\columnwidth]{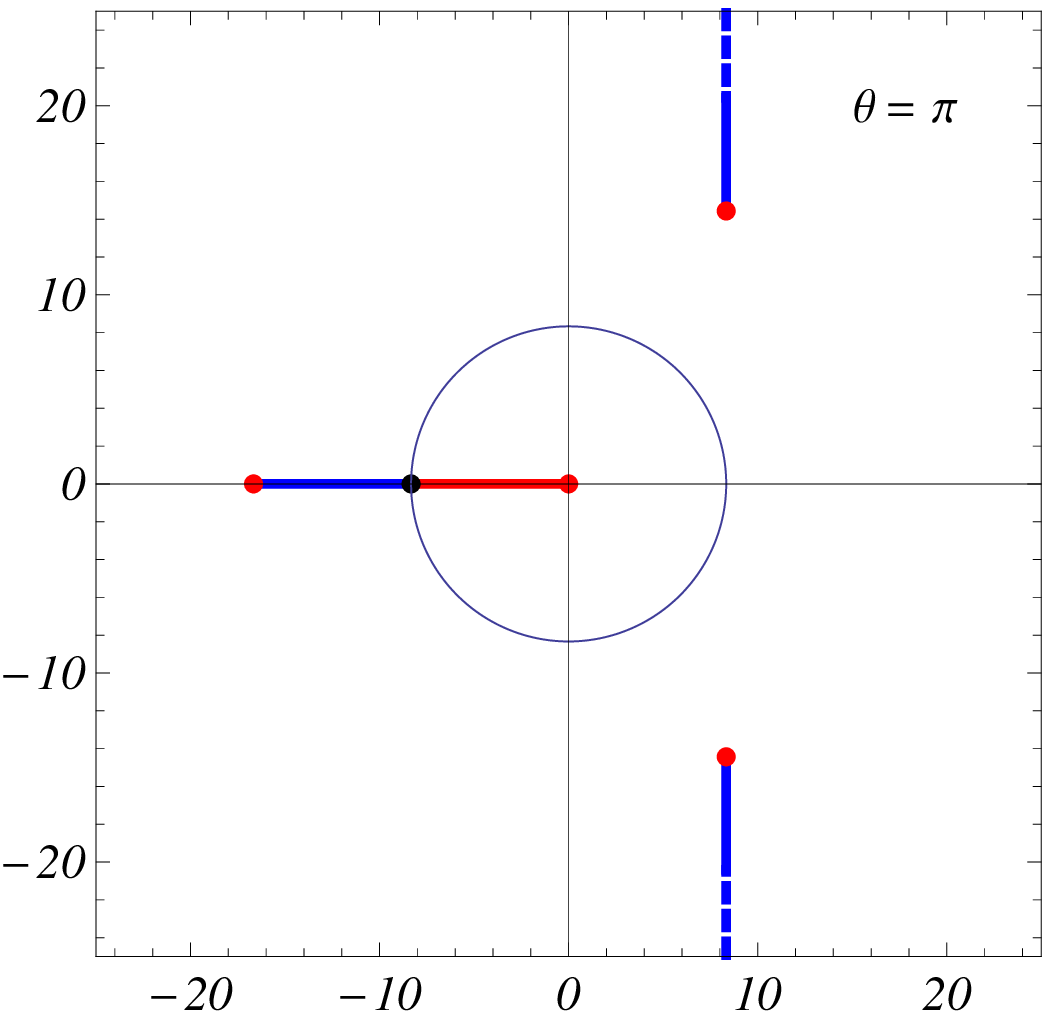}
	\caption{ (Color online) Motion of the Dirac points in momentum space for different values of the displacement angle $\theta$. The units are such that $c=1$, 					 $b=0.06$. When $\Delta=0$, the central Dirac point is surrounded by three Dirac points at distance $c/b$ (red dots). When increasing $\Delta$, the central Dirac 					 point merges with one of the three Dirac points, leaving the two remaining points isolated.  When varying $\theta$, the position of the merging point draws a circle 				of radius $c/2b$. The full curves represent the positions of the Dirac points until two of them merge. The dashed curves represent the position of the
	remaining Dirac points after merging of the other two. }
	\label{angular-dep}
	\end{figure}

\begin{figure}[h!]
\includegraphics[width=.7\columnwidth]{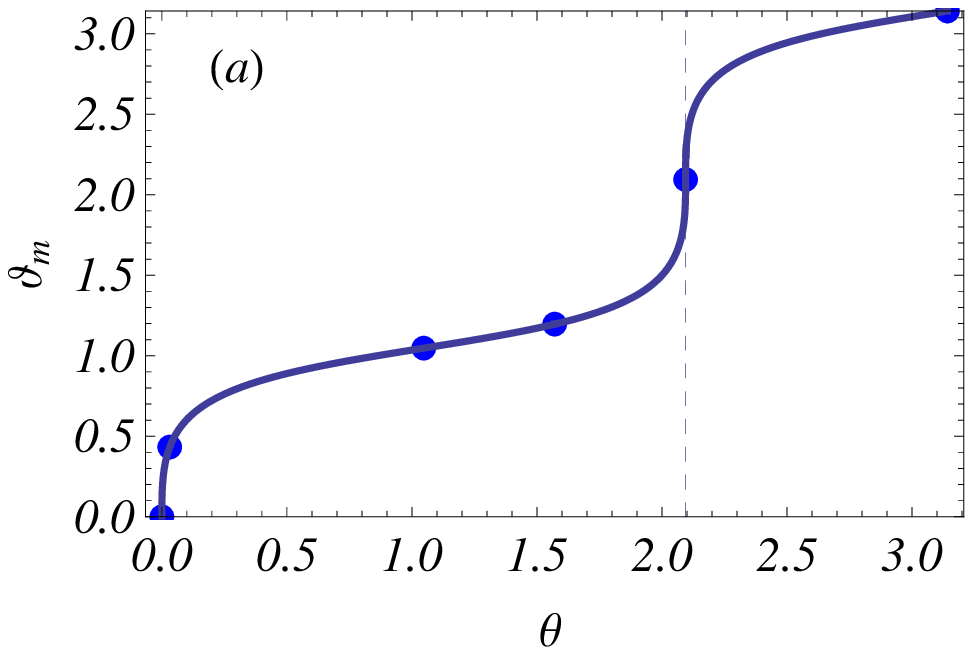}
	\includegraphics[width=.7\columnwidth]{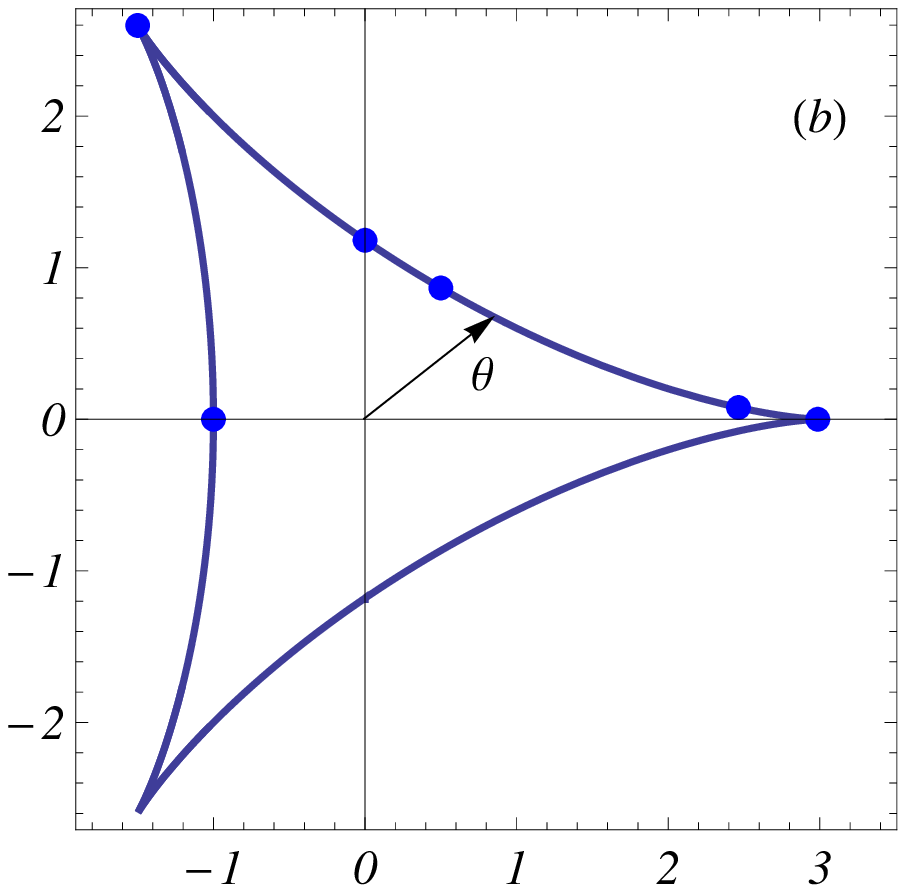}
	\caption{(Color online) (a) Dependence of the merging position angle $\vartheta_m$ as a function of the angle of deformation $\theta$.
The vertical dashed line for $\theta=2 \pi/3$ indicates the $2 \pi/3$ rotational symmetry.
(b) Polar plot of the angular dependence of the critical value  $\Delta_m(\theta)$ of the deformation at the merging.
$\Delta_m$ is given in units of $c^2/4b$. }
	\label{thetamp}
	\end{figure}

In addition to the above distortion along a high-symmetry axis of the lattice, we consider the more general deformation
along an arbitrary axis which corresponds to a complex-valued $\Delta$.
Fig. \ref{angular-dep} shows the evolution of the position of the Dirac points when increasing $\Delta$ for different values of the angle $\theta$ defined as $\Delta = |\Delta| e^{i \theta}$.
The complex position of the merging point in reciprocal space is
	\beq
	\pi_m(\theta) = {c \over 2 b} e^{i \vartheta_m(\theta)}
	\eeq
where the angular dependence of the merging angle $\vartheta_m(\theta)$ is given by
	\beq
	\label{theta}
	\tan \theta= { 2 \sin \vartheta_m - \sin 2 \vartheta_m \over \cos 2 \vartheta_m + 2 \cos \vartheta_m},
	\eeq
and is plotted in Fig. \ref{thetamp}(a).
For a given angle $\theta$, the merging is reached for a critical value  $\Delta_m(\theta)$ given by
\beq  \Delta_m(\theta)=  {c^2 \over 4 b} [5 + 4 \cos 3 \vartheta_m(\theta) ]^{1/2}, \eeq
which is shown in Fig. \ref{thetamp}(b).

Eq. (\ref{theta}) and Fig. \ref{thetamp}(a) reveal that the angular dependence of the merging point, $\vartheta_m$, is not linear in the angle of the deformation axis, $\theta$.
This is best captured in the vicinity of $\theta= 0$ or $2\pi/3$ where the slope of the Fig. \ref{thetamp}(a) increases abruptly.
Most saliently, a slight deviation from a high-symmetry axis ($\theta=0,\pm 2\pi/3$), i.e. an infinitesimal imaginary contribution
to the shift $\Delta$, renders the triple-merging point unstable. Indeed, as one may see from Fig. \ref{angular-dep}, only two Dirac points 
merge, whereas the third one remains isolated.
From this perspective, one can qualify the triple-merging scenario as unstable.
However, one can argue that for moderate residual chemical doping or in the presence of disorder, the difference between a triple-merging points and a single-merging point with a close-by extra Dirac cone is smeared out, such that the study of the triple-merging scenario
may still provide physical insight.

\subsection{Winding numbers}\label{topo:zero}

In Sec. \ref{sec:BandStruc}, we have encountered different merging types of Dirac points that fall into two classes: whereas
the merging transition is associated with the opening of a band gap, there are transitions, such as triple merging or those
encountered in twisted bilayer graphene,\cite{degail11} that are not accompanied by a gap opening. The nature of the different merging
transitions turns out to be determined by the underlying topological properties of the band Hamiltonian. In this section we discuss
these merging transitions in terms of winding numbers that play the role of topological charges the sum of which is conserved
across the transitions. Furthermore, these winding numbers play an eminent role also in the presence of a magnetic field, where
they determine the number of zero-energy modes and where
they intervene in the semi-classical treatment that describes to great accuracy the LL spectrum obtained from the full solution of
the quantum-mechanical equations (Sec. \ref{sec:topology}).

In the vicinity of band-contact points, the system may be described in terms of the effective two-band Hamiltonian
	\beq
	\mathcal{H}(\bq) =
	\begin{pmatrix}
	0 & h_x(\bq) - i h_y (\bq) \\
	 h_x(\bq) + i h_y (\bq) & 0
	\end{pmatrix},
	\eeq
diagonalization of which yields the energy spectrum $\epsilon_{\lambda}(\bq)=\lambda\sqrt{h_x^2(\bq) + h_y^2(\bq)}$
and the eigenstates
\beq \label{eq:WF}
\psi = {1 \over \sqrt{2} } \left(
                                 \begin{array}{c}
                                     1   \\
                                  \lambda e^{ i \phi_\bq}  \\
                                 \end{array}
                               \right) \eeq
where $\tan\phi_\bq=h_y(\bq)/h_x(\bq)$, and $\lambda = \pm 1$ denotes the band index. The relative phase $\phi_\bq$ exhibits a
particular topological structure that we
discuss in terms of the pseudospin map, which is defined as
	\beq\label{eq:map1}
	h :  \bq \longrightarrow \{ h_x (\bq), h_y (\bq) \}.
	\eeq
Because of the single-valuedness of the wave functions (\ref{eq:WF}), the map $h=[h_x(\bq),h_y(\bq)]$ must retrieve its original
value, modulo $2\pi$ on a closed path that starts and terminates on a precise value $\bq_0$. All closed paths therefore fall into
distinct homotopy classes that are described by the integer $w_\mathcal{C}$, which is an element of the homotopy group $\pi_1(S^1)$
associated with the map $h$ from the closed path $\mathcal{C}$ (with the topology of a circle $S^1$) in reciprocal space to closed
paths in pseudospin space. In order to calculate this integer, which is the \textit{pseudospin winding number},
one needs to integrate the Berry connection
$\mathcal{A}_\bq=i\psi^{\dagger}\nabla_{\bq}\psi$ over the closed path, in terms of the wave functions (\ref{eq:WF}) and
the reciprocal-space gradient $\nabla_\bq=(\partial/\partial_{q_x},\partial/\partial_{q_y})$. One obtains
\beq \label{eq:topoCp}
w(\mathcal{C})= \frac{1}{2 \pi} \oint_{\mathcal{C}}  \nabla_\bq \phi_\bq \cdot d\bq .
\eeq

As such, $w(\mathcal{C})$ is nothing other than the Berry phase \cite{Berry,falkovsky}
within a factor $\pi$ calculated over the path $\mathcal{C}$. However, we avoid the name ``Berry
phase'' in the present context for two reasons. First, the Berry phase does not necessarily need to be an integer, as it has been
shown e.g. in the case of gapped graphene (or boron-nitride) where the Berry phase explicitly depends on the energy of the
path.\cite{Fuchs10} Only the topological part, which should then be viewed as the winding number, of this Berry phase determines
the chiral properties, such as those revealed by the LL spectrum in the semi-classical approach discussed below. Second,
a quantum-mechanical phase is defined modulo $2\pi$, and one would therefore not expect different physical properties for
$\pi w_\mathcal{C}$ as compared to 0 for even values of $w_\mathcal{C}$ or $\pi$ for odd values.\cite{Park,falkovsky12} However, relevant
properties of the level spectrum, such as the dispersion relation at intermediate energies\cite{MF06,Park},
the degeneracy of the zero-energy modes, and their protection,
depend sensitively on the precise value of $w_\mathcal{C}$. Notice that $w_\mathcal{C}$ is an additive quantity -- if one devides
the surface $\Sigma$ enclosed by the path $\mathcal{C}$ into distinct pieces, $\Sigma_1 ... \Sigma_N$, the winding number is the sum
of the partial ones calculated over paths $\mathcal{C}_j$ encircling the surfaces $\Sigma_j$,
\beq
w\left(\mathcal{C}\right)=\sum_{j=1}^N w\left(\mathcal{C}_j\right).
\eeq
Furthermore, if one of the merging transitions discussed in Sec. \ref{sec:BandStruc} takes place inside a path $\mathcal{C}$, the
winding number is a conserved quantity. It is simply the sum of the winding numbers calculated on paths around the original band-contact
points before the merging transition and may thus also be viewed as a topological charge.

\begin{figure}[h!]
	\includegraphics[width = 0.45\columnwidth]{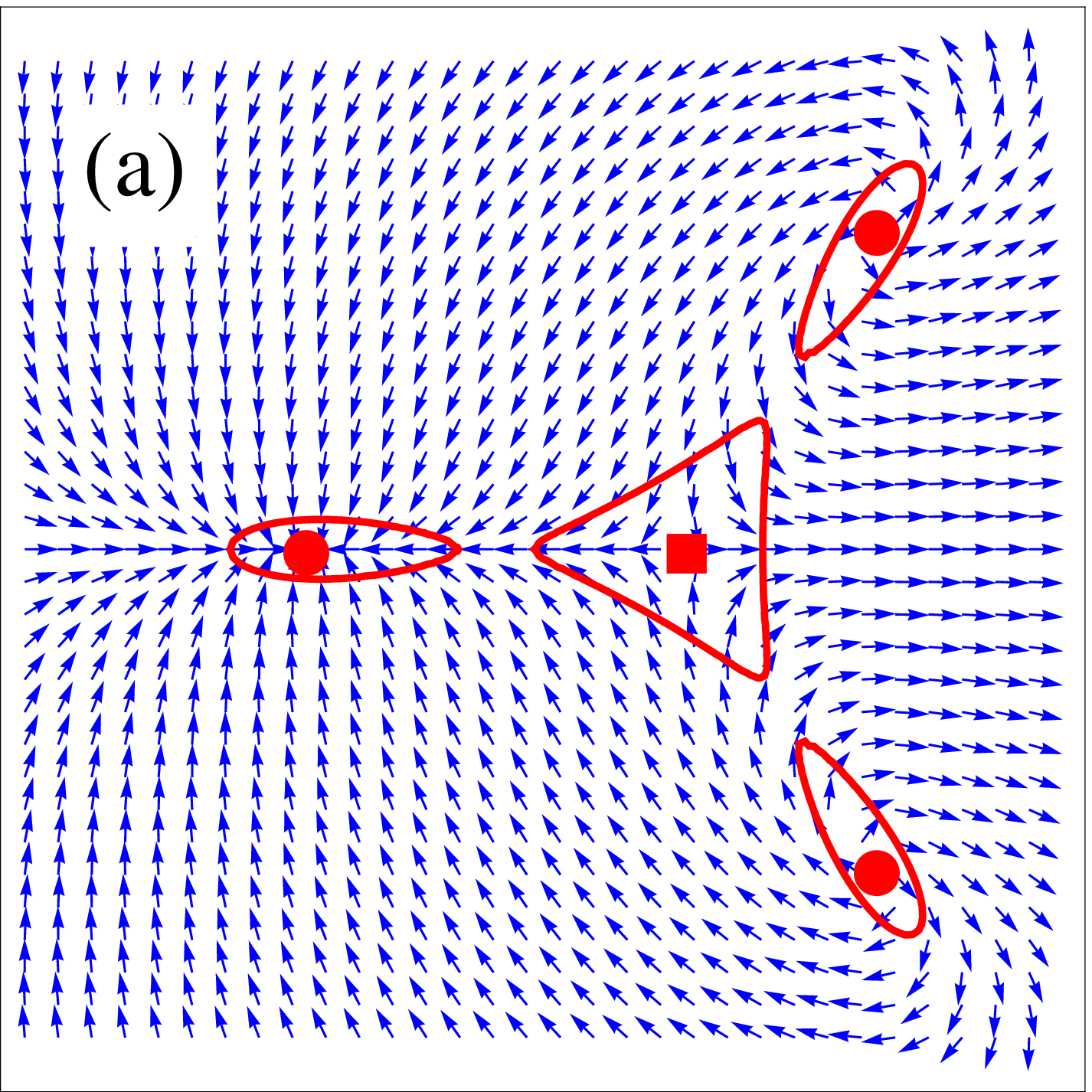}
\includegraphics[width = 0.45\columnwidth]{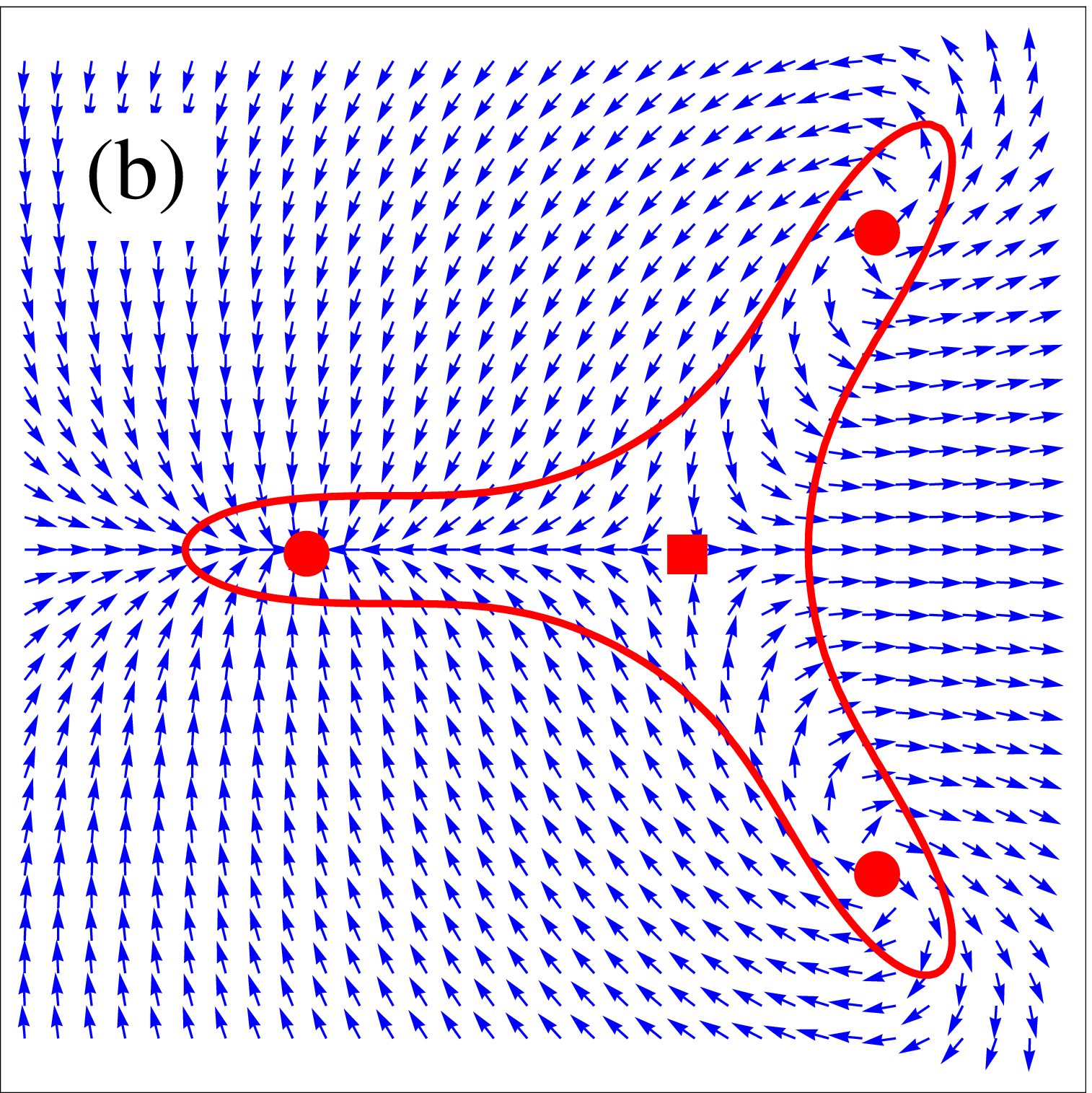}
\includegraphics[width = 0.45\columnwidth]{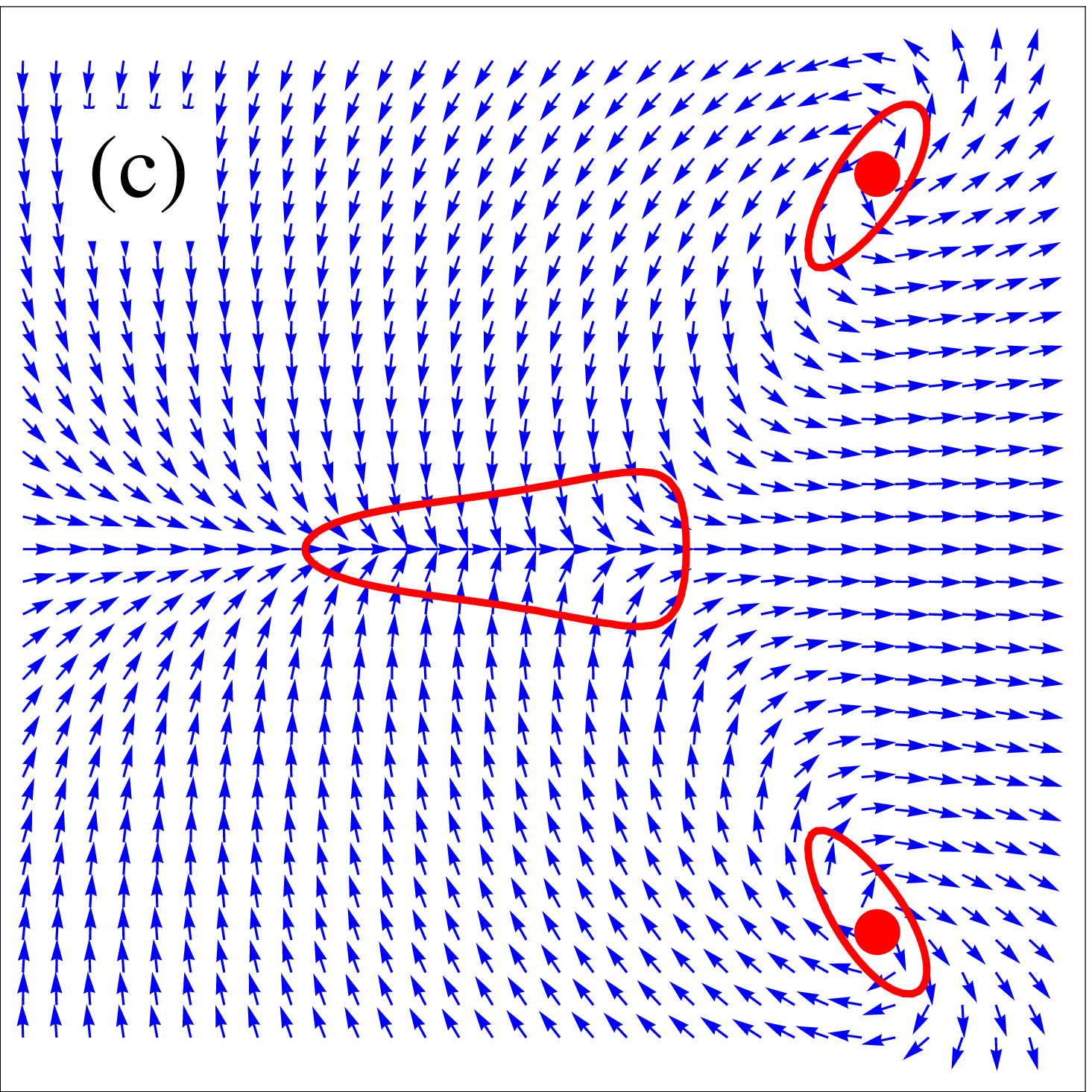}
\includegraphics[width = 0.45\columnwidth]{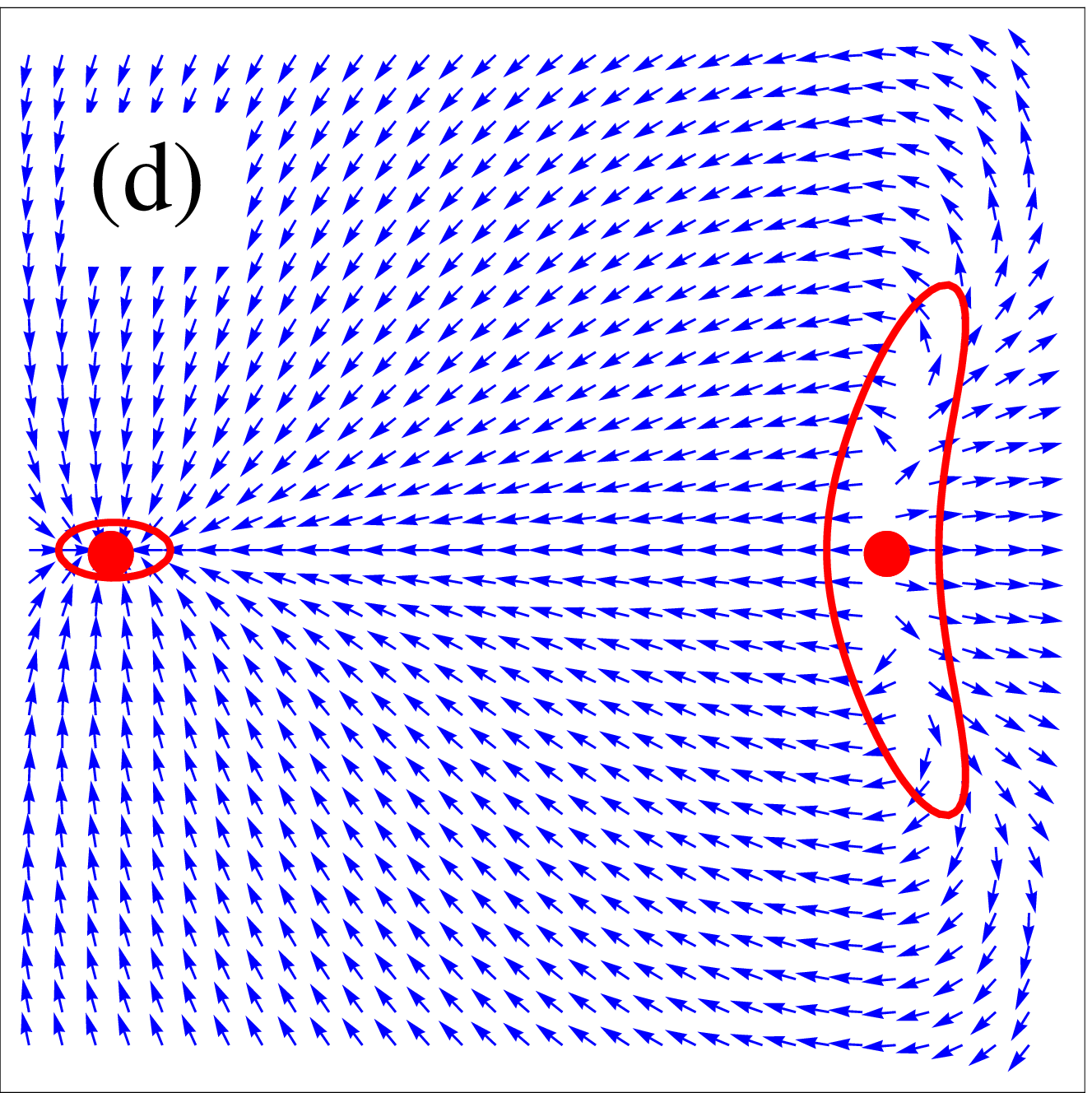}
	\caption{(Color online) Winding numbers for the pseudospin map.
The square indicates a charge  $-1$ and the circles indicate a charge $+1$. (a) In the sector L, all Dirac points are resolved and described by individual
topological charges. (b) Sector H, the possible closed loops enclose all Dirac points, and the topological charge is thus $2$. (c) Sector M, whereas
the Dirac points $B$ and $C$ are resolved (charge 1), the points $A$ and $D$ have merged, such that closed loops yield a charge 0. (d) Sector
M$^{\prime}$, the triple merging envolves the points $B$, $C$, and $D$ and closed loops yield a charge 1, in addition to the charge 1 stemming from
the isolated Dirac point $A$.}
	\label{winding}
	\end{figure}

As an example, we plot the map (\ref{eq:map1}) in Fig. \ref{winding} for the Hamiltonian $\mathcal{H}_b+\mathcal{H}_c+\mathcal{H}_{\Delta}$
in different configurations corresponding to the sectors L, H, M and M' of Fig. \ref{fig:secteurs}. In the low-energy sector (L),
  for energies below both saddle points $E_S$ and $E_{S'}$, all
Dirac points are resolved, and there exist thus closed loops encircling each of the points [Fig. \ref{winding}(a)]. The vicinity of the
points $A$, $B$, and $C$ is then described by a charge $+1$ each, whereas the central point $D$ carries a charge $-1$. At energies larger
than $E_S$ and $E_{S'}$
[sector H, Fig. \ref{winding}(b)], all closed loops necessarily enclose all points, and the topological charge is therefore the sum ($+2$)
of all individual Dirac points resolved at low energies. This situation is to be contrasted to the sector M, for energies $E$ with $E_S<E<E_{S'}$,
[Fig. \ref{winding}(c)]. The points $A$ and $D$
are then necessarily enclosed by all corresponding loops, such that the charge is 0, whereas a second class of loops can still resolve
the points $B$ and $C$ (charge $+1$ each). In the sector M', for $E_{S'}<E<E_{S}$, the three points $B$, $C$, and $D$ can no longer be resolved (loops of
charge $+1$), whereas $A$ remains a Dirac point with charge $+1$ [Fig. \ref{winding}(d)].

In view of the different merging transitions, we have already mentioned that the pseudospin winding number is conserved during such
transitions. For merging ($\Delta<0$) of two Dirac points described by winding numbers of opposite sign (e.g. $w_D=-1$ and $w_A=+1$)
-- this is necessarily
the case for Dirac points that are related by time-reversal symmetry -- the topological charges are thus annihilated across the
transition, such that the zero-energy states are no longer topologically protected. One therefore observes the opening of a local band
gap that is associated with the merging of Dirac points with opposite winding numbers [see Figs. \ref{band}(a) and (b)]. In
the case of a triple merging ($\Delta>0$), the sum of the winding numbers is $w_B+w_C+w_D=+1$, such that any path enclosing the point
where the Dirac points $B$, $C$, and $D$ have merged carries a winding number $+1$ also after the transition. The (zero-energy)
band-contact is therefore preserved and the opening of a band gap topologically prohibited [see Figs. \ref{band}(c) and (d)].

	\section{Landau Level Spectrum}\label{sec:LL}
		\begin{figure}[h!]
		\includegraphics[width=.9\columnwidth]{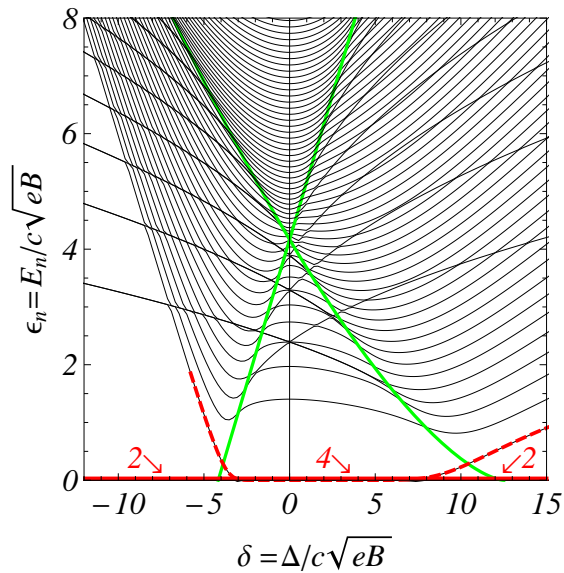}
		\caption{(Color online) LL spectrum of the Hamiltonian (\ref{MagneticHam}) as a function of $\delta$. The parameter $\beta$
is fixed to $0.06$.
		The spectrum is cut into four parts delimited by the thick (green) lines which depict the energy of the saddle points
in units of $c/l_B=c\sqrt{eB}$. The figure shows the existence of four zero-energy modes. Two of them (red) are topologically stable.
The other two (here only the one of positive energy is marked with a dashed curve) acquire a finite energy for sufficiently large values of $\delta$.}
		\label{spectrum}
		\end{figure}
The considerations of the previous section on the band structure in the absence of a magnetic field yield valuable insight into
the LL spectrum, which is formed when a perpendicular magnetic field, $B\be_z=\nabla\times \mathbf{A}$,
is applied to the graphene layers. In this section, we compare the LL spectrum obtained from a numerical solution of the
full quantum-mechanical problem described by the Hamiltonian $\maH_b + \maH_c + \maH_\Delta$ in the presence of a magnetic field
(Sec. \ref{sec:LQ}) to that calculated within a semiclassical approximation (Sec. \ref{sec:semicl}).
  A detailed discussion of the LLs in the different energy sectors (L, M, M', and H) is postponed to Sec. \ref{sec:discussion}.

		\subsection{Landau Quantization}\label{sec:LQ}
The magnetic field may be taken into account with the help of the Peierls substitution (for electrons of charge $-e$)
\beq\label{eq:Peierls}
\Pi = \pi + e \mathbf{A},
\eeq
where $\mathbf{A}$ is the vector potential.	
This allows one to introduce the harmonic oscillator operators
\beq\label{eq:ladder}
a = \frac{l_B}{\sqrt{2}} \left( \Pi_x - i \Pi_y \right), \qquad a^{\dagger} = \frac{l_B}{\sqrt{2}} \left( \Pi_x + i \Pi_y \right),
\eeq
with $[a,a^{\dagger}] = \mathbbm{1}$.
The magnetic length $l_B = 1/\sqrt{eB}\simeq 26\, \text{nm}/\sqrt{B\text{[T]}}$ encodes the size of the cyclotron orbits in real space.

In the presence of a magnetic field, the Hamiltonian $\maH_b + \maH_c + \maH_\Delta$ reads
\beq
\label{MagneticHam}
\frac{\mathcal{H_B}}{c/l_B} =
\left( \begin{array}{cc} 0 & 2\beta a^2 - \sqrt{2} a^{\dagger} - \delta \\ 2\beta a^{\dagger2} - \sqrt{2} a - \delta & 0 \end{array} \right).
\eeq
We have rescaled the energy with respect to $c/l_B=c\sqrt{eB}$,
the characteristic LL energy of the central Dirac cone $D$ for $\delta = 0$, and have also introduced the dimensionless ($B$-field-dependent) shift $\delta = \Delta/c\sqrt{eB}$ as well as the parameter
\beq\label{def:beta}
\beta = \frac{b\sqrt{eB}}{c}.
\eeq
The quantity $\beta$, which measures the amplitude of the trigonal warping in units of the inverse magnetic length,
is a central parameter in the description of the LL spectrum. It may also
be interpreted as the inverse of the reciprocal-space
distance $c/b$ of a peripheral Dirac cone ($A,B,C$) to the central one ($D$) and the magnetic length $l_B$. Viewed as an energy scale,
it is proportional to the ratio between  the first excited LL ($c \sqrt{2 e B}$) and the energy $E_S$ of the saddle points joining the cones 
[in the absence of a deformation (at $\delta = 0$)]. From (\ref{saddle-point}), we have

\beq\label{eq:crit1}
\frac{c\sqrt{2 e B}}{E_S}= 4 \sqrt{2} \beta.
\eeq
Finally, the
parameter $\beta$ turns out to describe the role of magnetic blurring that is described in Sec. \ref{sec:topologyB}.
Notice that one might also have performed the Peierls substitution in the original four-band model (\ref{Hamlattice}), as it has been done 
for the case without trigonal warping.\cite{peeters} However, the corrections are weak in the low-energy limit that we are interested in, and the 
effective two-band model (\ref{MagneticHam}) provides a good description of the LL spectrum.

The numerically obtained spectrum of Hamiltonian (\ref{MagneticHam}) is plotted in Fig. \ref{spectrum} as a function of $\delta$, which corresponds to varying $\Delta$ and/or $B$ as well as the energy of the saddle points in order to sustain the same number of LL below $E_S$.
Notice that we have only plotted the spectrum at positive energy, $\epsilon_n$, those at negative
energy are obtained from the plotted ones simply by adding a minus sign, $-\epsilon_n$, as a consequence of the particle-hole symmetry
respected by Hamiltonian (\ref{MagneticHam}).
Increasing the value of $\beta$ will then only scroll the levels up in energy and reduce the number of modes within the trigonally warped area.
For this reason, we focus on an arbitrarily low value of $\beta = 0.06$.
The spectra for other values of $\beta$ are discussed in Sec. \ref{sec:topologyB}

		\subsection{Semiclassical description}\label{sec:semicl}
In order to reproduce the spectrum of Fig. \ref{spectrum} and to understand the underlying physical properties, we rely on a semiclassical analysis.
 			\begin{figure}[h!]
			\includegraphics[width=.9\columnwidth]{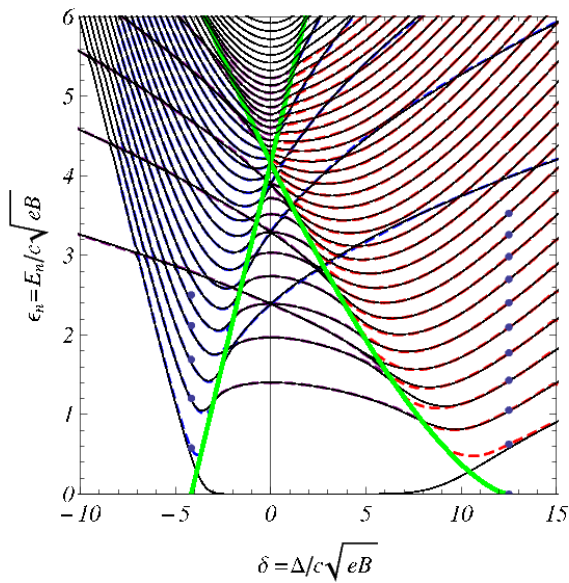}
			\caption{(Color online) Semiclassical reconstruction (dashed lines)
of the spectrum in Fig. \ref{spectrum}, for $\beta=0.06$. The different regions of the spectrum are discussed in detail in Sec. \ref{sec:discussion}.
The  blue  dots indicate the LL spectrum at the merging (left) and the triple-merging
transition (right), in which case the LLs scale as $(n+1/2)^{2/3}$ and $n^{3/4}$, respectively,
see Secs. \ref{sec:merging.transition} and \ref{sec:triple.merging}. The green lines indicate the energies of the saddle points $\ep_S$ and
$\ep_{S'}$. }
			\label{semiclassical}
			\end{figure}
This theory states, according to Onsager's argument,\cite{onsager,onsager2} that the reciprocal-space
area $\mathcal{A_C}(\epsilon_n)$ enclosed by the band contour
$\mathcal{C}$, for energy $\epsilon_n$, must fulfill
	\beq\label{eq:ons}
	\mathcal{A_C}(\epsilon_n) =\int_{\bk(\epsilon\leq \epsilon_n)} d^2k= 2\pi eB\left(n+ \gamma \right),
	\eeq
  where the mismatch factor
\beq
\gamma= \frac{1}{2}-\gamma_B \label{gammaM}
\eeq
has a contribution $1/2$ from the usual Maslov index for the harmonic oscillator and a second one, $\gamma_B$, that was first
identified with the Berry phase\cite{Wilkinson,mikitik} acquired on the path $\mathcal{C}$, whereas it has been shown afterwards
that only the topological part of the Berry phase enters into the expression.\cite{Fuchs10} Here, we express the quantity
$\gamma_B$ in terms of the pseudospin winding number, $\gamma_B=|w_\mathcal{C}|/2$, such that Eq. (\ref{gammaM}) becomes\cite{ecrys}
\beq
\gamma= \frac{1}{2}-\frac{|w_\mathcal{C}|}{2} \label{gammawC}.
\eeq
At first sight, the large-$n$ limit of the semiclassical approximation could be described in terms of a quantum-mechanical Berry phase
$\pi|w_\mathcal{C}|$ modulo $2\pi$, that is one identifies all odd and all even winding numbers, if one redefines the integer $n$.
However, Eq. (\ref{gammawC}) bears information about the presence and the number of zero-energy modes [see Sec. \ref{sec:topology}].

In order to obtain the semiclassical LL spectrum $\epsilon_n =\epsilon(n)$, we numerically invert Eq. (\ref{eq:ons}). The results
are shown in Fig. \ref{semiclassical} (dashed lines) in comparison with the ones (full lines)
obtained from a numerical solution of the quantum-mechanical eigenvalue equation [Hamiltonian (\ref{MagneticHam})].

\begin{figure}[h!]
\includegraphics[width=0.8\columnwidth]{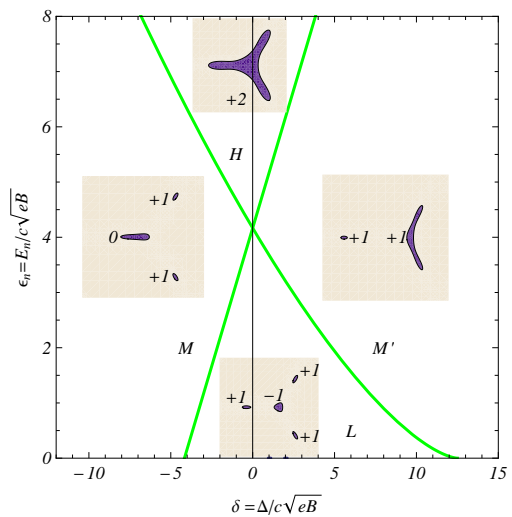}
\caption{(Color online) The different winding numbers attached to the different pockets imply different quantization rules.
}
\label{zonesberry}
\end{figure}

In addition to the LL spectrum, Figs. \ref{spectrum} and \ref{semiclassical} depict the energy of the saddle points rescaled by the energy $c/l_B$ 
(thick green lines),
\beq
\epsilon_S = E_S/c\sqrt{eB}, \qquad \epsilon_{S'} = E_{S'}/c\sqrt{eB}.
\eeq
Whenever a LL crosses one of the three saddle points, its properties, such as its degeneracy, are drastically modified due to the Lifshitz
transition involved.
Indeed, Figs. \ref{spectrum} and \ref{semiclassical} reveal four distinct sectors L, H, M, and M$^{\prime}$ introduced in Fig. \ref{fig:secteurs}
according to whether the energy is below both
saddle points [low-energy sector (L)], below only one of the saddle points [merging (M) and triple-merging (M$^{\prime}$) sectors],
or above both of them for the high-energy sector ($H$) [see Fig. \ref{zoneslandau}(a)].
For these distinct regions, Onsager's quantization reads differently, {because the winding number $w_\mathcal{C}$ takes different values for
differents types of orbits. The different sectors are shown on Fig. \ref{zonesberry} and discussed in detail in Sec. \ref{sec:discussion}.

\section{Zero modes and semiclassical quantization rule}\label{sec:topology}
 
 In principle, the semiclassical description is valid at large energies, whereas the zero modes require a specific (quantum-mechanical)
treatment. However, the semiclassical analysis and the intervening winding numbers provide valuable insight into the degeneracy of the zero modes.
The discussion of this relation is the issue of the present section. In Sec. \ref{sec:Rel0Modes}, we provide a simplified model to illustrate
this relation, whereas we discuss the LL degeneracy lifting due to magnetic blurring in Sec. \ref{sec:topologyB}.

\subsection{Relation between zero modes and winding number}\label{sec:Rel0Modes}

 We provide here a heuristic argument relating the total number of topologically protected zero-energy modes to the semiclassical quantization rule.
Consider first the model Hamiltonian  \cite{Volovik,Manes,ecrys}
\beq \label{eq:ModelHam}
{\mathcal H}= \Lambda \left(
                            \begin{array}{cc}
                              0 & \pi^{\dagger p} \\
                              \pi^{p} & 0 \\
                            \end{array}
                          \right)
                          \eeq
describing a band contact point with energy spectrum $\ep =\pm \, \Lambda |\bq|^{p}$ and with a winding number $p$. The global parameter $\Lambda$
has the physical dimension of an energy times the $p$-th power of a length.
In a magnetic field, performing the Peierls substitution (\ref{eq:Peierls}) and the replacement in terms of ladder operators (\ref{eq:ladder}),
one obtains the LL spectrum in a magnetic field,
\beq\label{eq:LLmodel}
\ep_n(B)= \pm \Lambda (2 e B)^{p/2} \sqrt{n(n-1) \cdots (n-p+1)},
\eeq
which, in the large-$n$ limit, may be approximated
as
\beq\label{eq:LLmodelSC}
\ep_n(B)\simeq \pm \Lambda \left[2 e B\left(n+{1 \over 2} - {p \over 2}\right)\right] ^{p/2}.
\eeq
This corresponds precisely to the semiclassical quantization
rule (\ref{eq:ons}) if we identify $p$ with the total winding number $w_{\mathcal C}$ in Eq.
(\ref{gammawC}). From Eq. (\ref{eq:LLmodel}), one notices that the $p$ quantum numbers $n=0,..., p-1$ correspond to states at zero energy.
Indeed, these states may be obtained from the eigenvalue equation
\beq
{\mathcal H}\left(
\begin{array}{c}
u_n\\ v_n
\end{array}\right)
=0,
\eeq
which is satisfied for the states
\beq
\psi_n^{(0)}=\left(
\begin{array}{c}
0 \\ |n\rangle
\end{array}\right), \qquad \text{for}\qquad n=0,...,p-1
\eeq
in terms of the eigenstates $|n\rangle$ of the number operator $a^{\dagger}a$, $a^{\dagger}a|n\rangle = n|n\rangle$. Because of the orthogonality
of the states, the zero-energy manifold is $p$-fold degenerate, i.e. the degeneracy corresponds to the total winding number $w_{\mathcal C}=p$,
as stated above.

The situation is different when there are more band-contact points, with different (local) winding numbers.
Consider a Hamiltonian describing $p$ massless Dirac points with winding number  $+1$, situated at the
complex positions $\alpha_i$ in reciprocal space, and $p'$ massless Dirac points with  winding  number $-1$, at the positions $\beta_j$. (Notice
that band contact points with larger winding numbers may be obtained by making several positions $\alpha_i$ or $\beta_j$ coincide.)
The Hamiltonian can be written as
\beq \mathcal{H}_\bq= \Lambda \left(
                \begin{array}{cc}
                  0 & f_\bq \\
                  f^*_\bq & 0 \\
                \end{array}
              \right), \label{GH}  \eeq
with $f_\bq=   \prod_{j=1}^{p'} (\pi^{\dagger} - \beta_j^*) \prod_{i=1}^{p} (\pi - \alpha_i)$ and $\Lambda$ is a global constant of the dimension
energy times the $(p+p')$-th power of a length.
The total number of Dirac points, and thus, after the Peierls substitution (\ref{eq:Peierls}),
the maximal number of zero-energy LL,  is $w_t= p+p'$. However, this $(p+p')$-fold degeneracy of the zero modes may be
partially lifted upon merging of two or more Dirac points.
In order to find the total number of topologically protected zero-energy levels, we thus continuously modify the parameters
\beq\label{eq:TotalMerg}
\alpha_i\rightarrow 0 \qquad \text{and}\qquad \beta_j\rightarrow 0,
\eeq
so that $f_\bq$ becomes $ {\pi^\dagger}^{p'} \pi^{p}$.  In a magnetic field, assuming for example that $p >p'$, this term is of the
form $\sqrt{2 e B}^{w_t} (a^\dagger a)^{p'} a^{p-p'}$, and the associated LL spectrum reads
\beq\label{eq:LLmodelBis}
\ep_n(B)= \pm \Lambda (2 e B)^{w_t/2}n^{p'}\sqrt{n(n-1) \cdots (n- p+p'+1)}.
\eeq
The same arguments as those presented in the discussion of the Hamiltonian (\ref{eq:ModelHam})
indicate that there are $w_p=p-p'$ zero-energy levels that correspond to the quantum numbers $n=0, ..., p-p'-1$
. Moreover, the same large-$n$ expansion as in the case discussed above yields
the spectrum
\beq\label{eq:LLmodelSCBis}
\ep_n(B)\simeq \pm \Lambda (2 eB)^{\frac{p+p'}{2}}n^{p'}
\left(n+{1 \over 2} - {w_p \over 2}\right) ^{w_p/2},
\eeq
which may be cast into the semiclassical quantization rule with a winding number $w_{\mathcal C}=w_p=p-p'$. Again the total winding number
is identical to the number of zero-energy modes.

These arguments show that,
although the maximum number of zero-energy LLs is
$w_t=p+p'$, a quantum-mechanical coupling between them partially lifts the degeneracy, but $w_p=|p-p'|$ zero modes remain
topologically protected. Applied to the model (\ref{fullHam}), one has $p=3$ and $p'=1$, so that the maximal number of zero modes
is $w_t=4$ and the number of topologically protected modes is $w_p=2$.

		\subsection{Magnetic blurring}\label{sec:topologyB}

It is apparent from Fig. \ref{semiclassical} that the semiclassical treatment (\ref{eq:ons}) based on Onsager's quantization
rule provides a reliable description of the LL spectrum in the major part of the parameter range. However, it is challenged
in the vicinity of the saddle points $\epsilon_S$ and $\epsilon_{S'}$.
Intuitively, one may understand the failure
of semiclassical quantization if one considers the topological winding number (\ref{eq:topoCp})
that is calculated from closed loops around the remarkable
points in reciprocal space [see Fig. \ref{winding}] -- in the presence of a strong magnetic field, these loops are at odds with quantum
mechanics because the components of the wave vector are no longer good quantum numbers, such that the images of the loops defined by the
maps (\ref{eq:map1}) are constrained by a Heisenberg uncertainty relation.

Indeed, in the presence of a magnetic field, the momenta no longer verify the simple commutation relation $[\Pi_x,\Pi_y] = 0$ but rather a Heisenberg algebra
\beq\label{eq:CommRel}
[\Pi_x,\Pi_y] = -\frac{i}{l_B^2}.
\eeq
An immediate consequence of this non-commutative geometry is that reciprocal space is now ``patched'' or ``blurred'' by irreducible regions of area $1/l_B^2$ below which it is impossible to resolve the physical properties of electrons in a magnetic field.
This is similar to the phase space of a one-dimensional quantum-mechanical particle, which is devided into minimal regions of the
size of the Planck constant $h=2\pi$, below which the physical properties of the particle cannot be resolved.
As a consequence, the winding of the pseudospin vector cannot be determined by paths the area of which encloses less than the minimal area of $\propto 1/l_B^2$, which plays the role of the Planck constant in reciprocal space.
If we were to define a winding number in the non-commutative reciprocal space, we should then consider larger and larger contours as the 
magnetic field increases since $1/l_B^2 \propto B$, as shown in Fig. \ref{map}.
	\begin{figure}[h!]
	\includegraphics[width=.49\columnwidth]{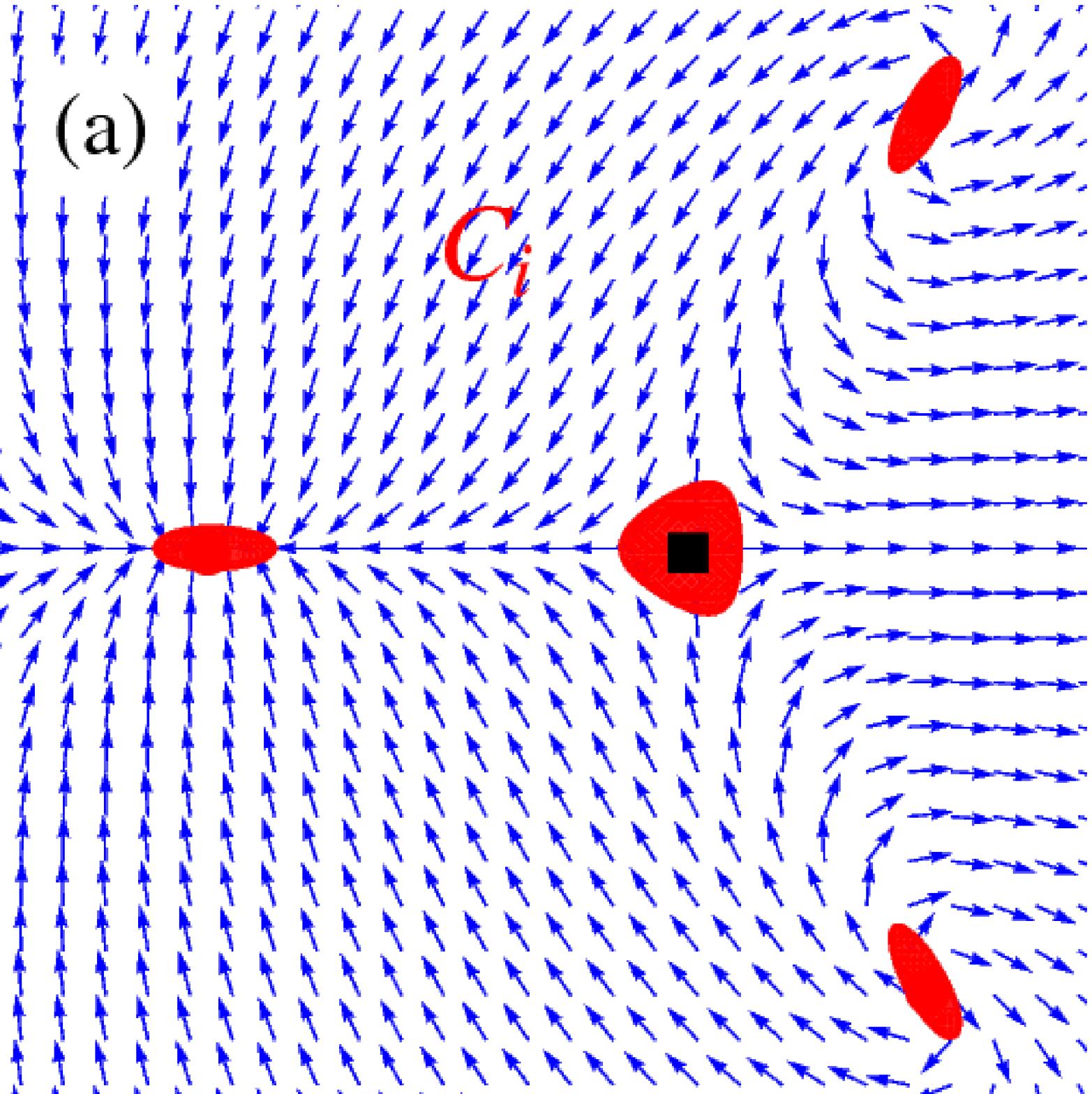}
	\includegraphics[width=.49\columnwidth]{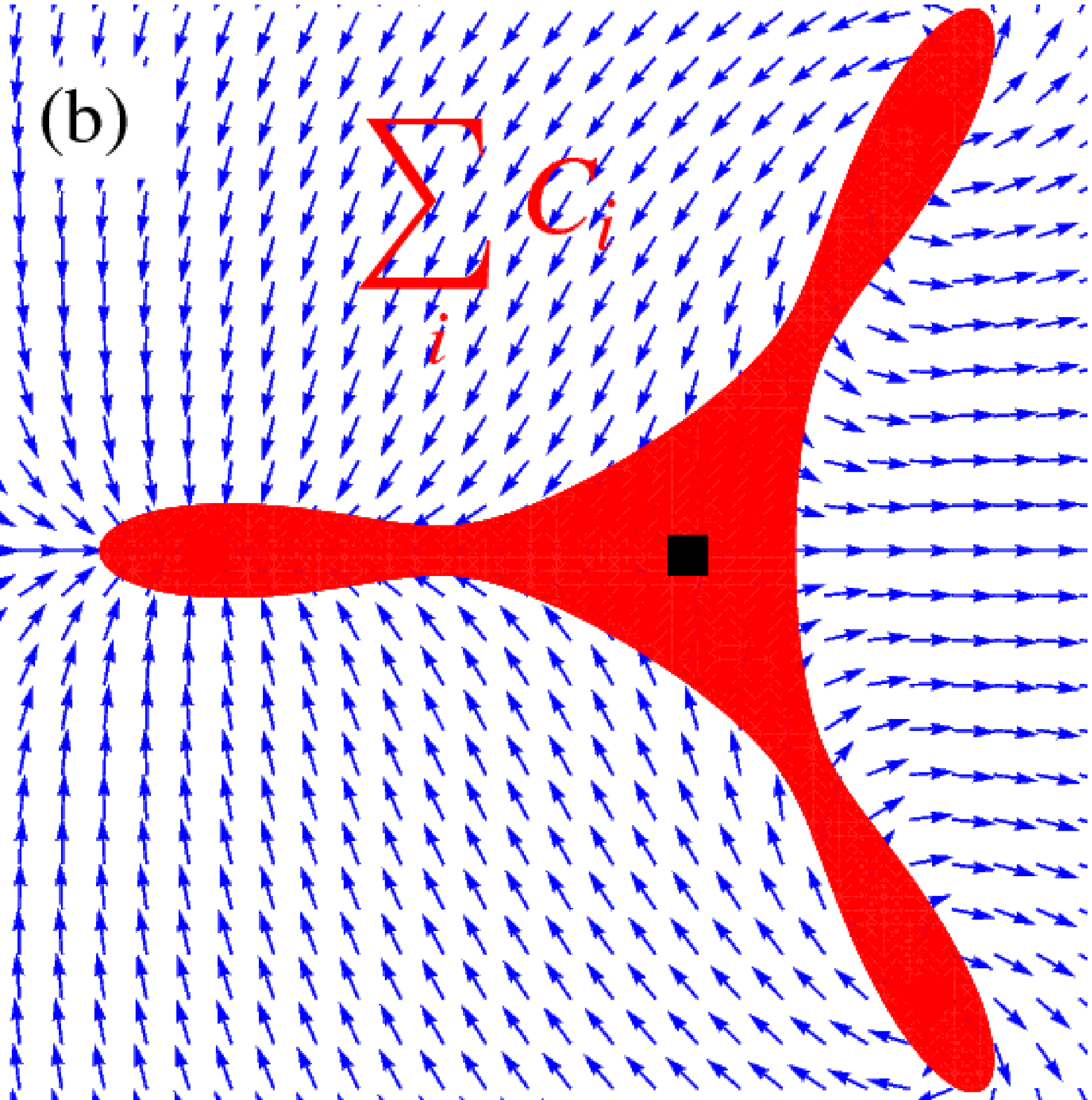}
	\caption{(Color online) Magnetic blurring for zero modes. Each contour encloses a minimal surface of $\sim 1/l_B^2=eB$ in reciprocal space 
(red areas). 
(a) At low magnetic fields, the blurring is low, and each minimal surface contains a single Dirac point.
The semiclassical quantization rule holds for each contour encircling a Dirac point. 
(b) When the field increases, the contour associated with the zero mode encloses a larger surface so that the individual Dirac points are
no longer resolved.
The effective winding number experienced by the electron is $w_p=\sum_i w_{i} $. 
}
	\label{map}
	\end{figure}

One may thus pictorially understand that whenever the irreducible area of $1/l_B^2$ becomes too large,
the   relevant winding contours enclose inevitably   more than one singularity. This is shown in Fig. \ref{map} where at low fields
the contours $\mathcal{C}_i$ around the individual singularities enclose each a winding number $w_i= \pm 1$ whereas at high fields 
the blurred contour encloses
a winding number  $\sum_i w_{i}= 2$.
Therefore, for a sufficiently strong magnetic field, the only relevant quantity is the total winding
number around \textit{all} the singularities,
	\beq\label{winding_lowB}
	w_p = \left|\sum_i w_i\right|,
	\eeq
as opposed to the total sum of the winding numbers
	\beq\label{winding_highB}
	w_t = \sum_i \left|w_i\right|.
	\eeq
which sets the total number (but not necessarily protected) of Dirac points. In the model Hamiltonian discussed in Sec. \ref{sec:Rel0Modes},
this magnetic blurring may be viewed alternatively as an effective merging of the band-contact points (\ref{eq:TotalMerg}).

As a consequence of the above arguments, increasing the magnetic field induces, even at zero energy, a Lifshitz transition that is
characterized by a partial degeneracy lifting of the zero-energy LL from $w_t$ (per spin and valley) to $w_p\leq w_t$, while $w_t-w_p$
levels disperse as a function of $B$.


	\begin{figure}[h!]
		\includegraphics[width=.49\columnwidth]{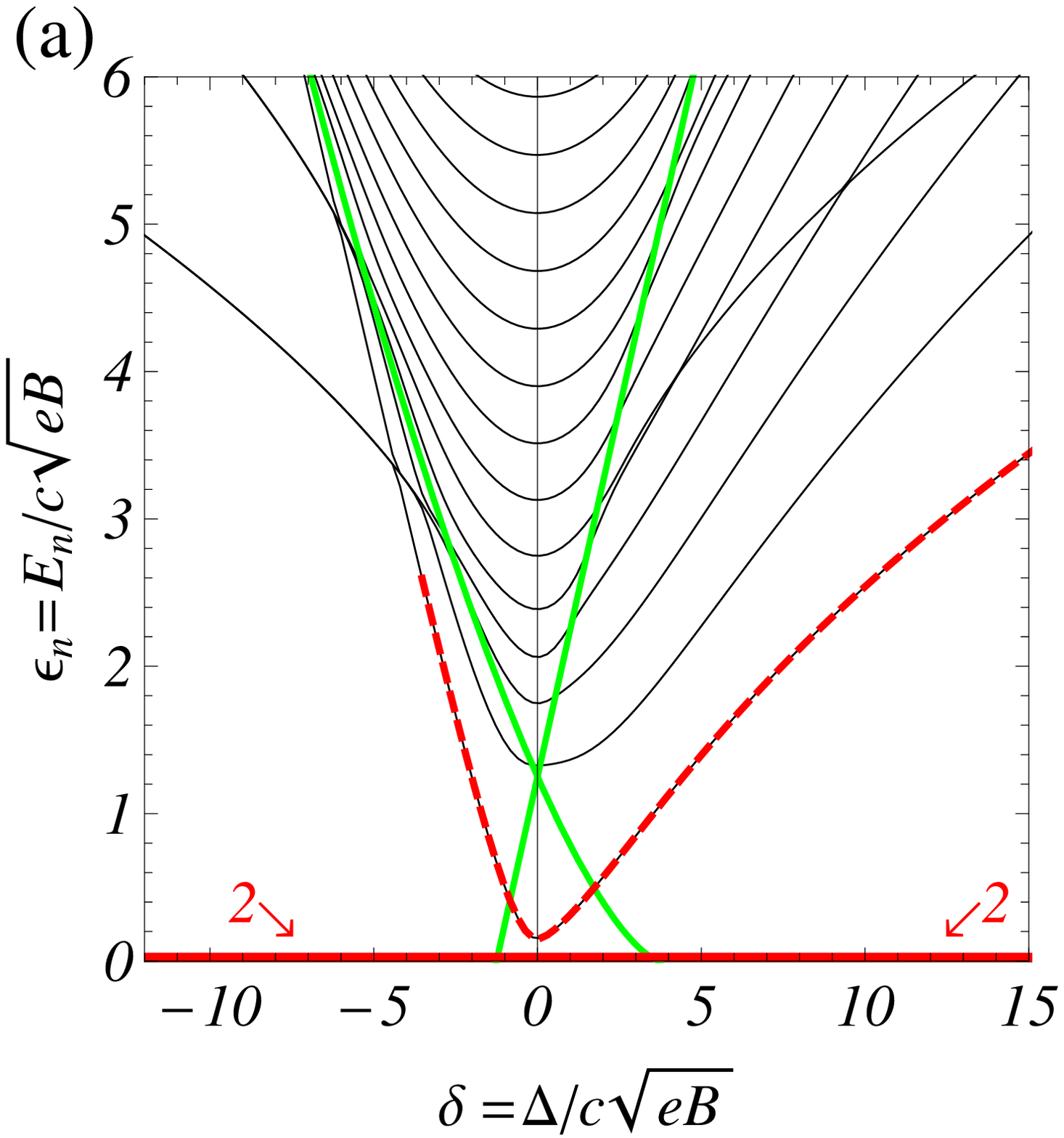}
	\includegraphics[width=.49\columnwidth]{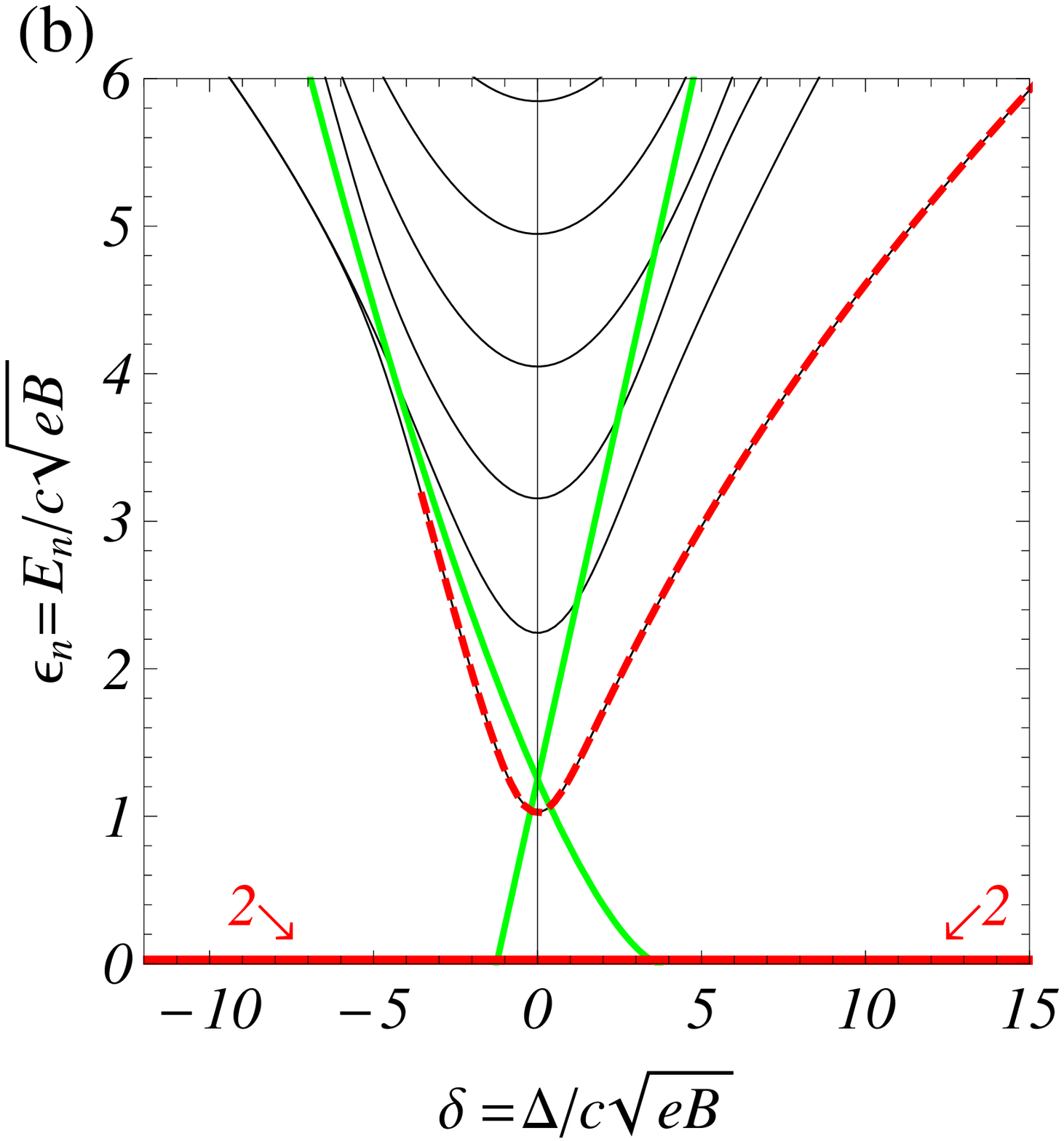}
				\includegraphics[width=0.9\columnwidth]{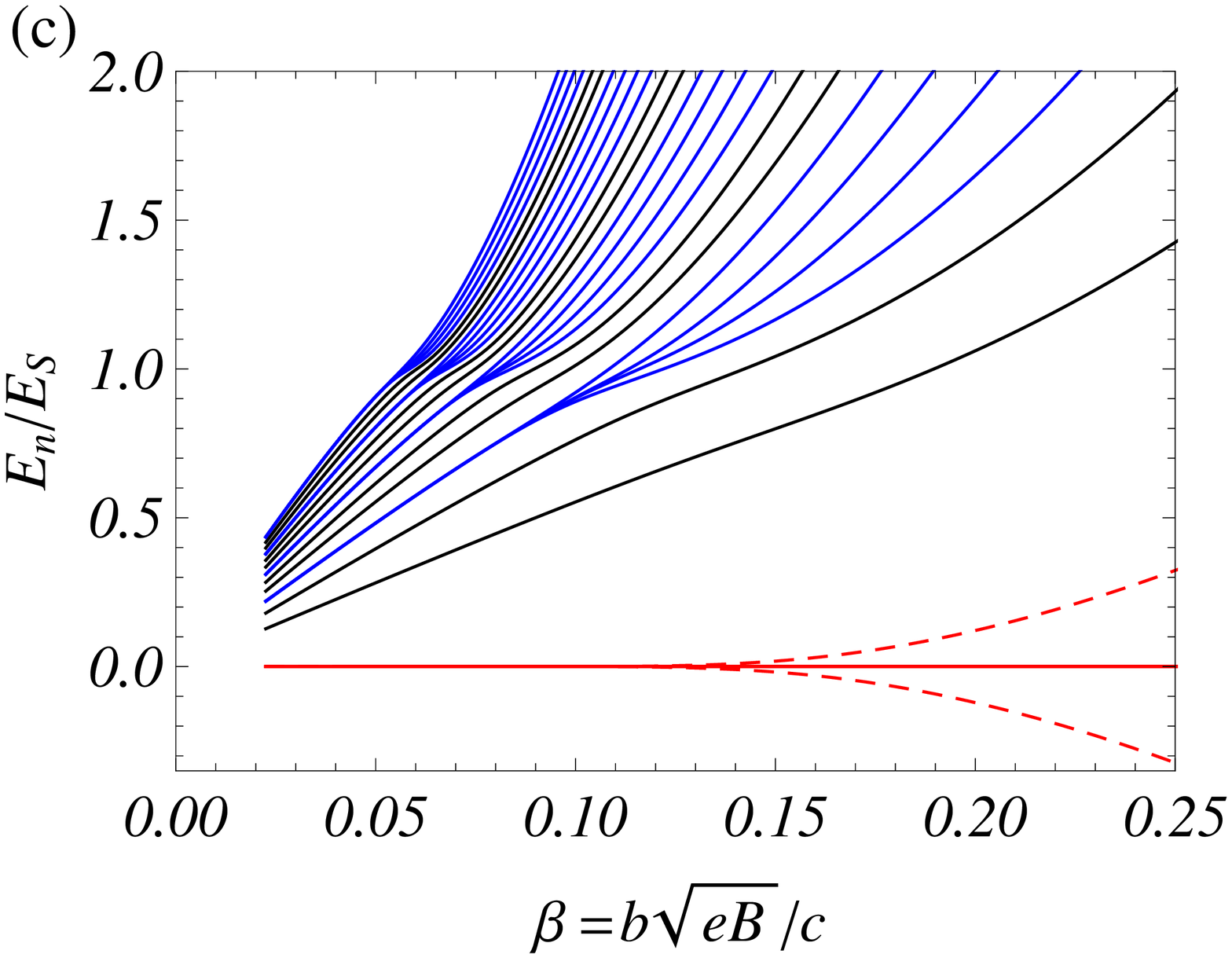}
		\caption{(Color online) LL spectrum of the Hamiltonian (\ref{MagneticHam}) as a function of $\delta$. The parameter $\beta$ is fixed
to $0.2$ (a) and $0.45$ (b). (c) LLs as a function of the parameter $\beta$ in the absence of distortion ($\delta=0$). 
It exhibits the lift of the fourfold degeneracy when the energy of the LLs becomes larger than the saddle point energy $E_S$ (blue levels). 
Two zero energy levels stay stable while two other levels get a finite energy when $\beta \gtrsim 1/4 \sqrt{2} \simeq 0.18$.
		}
		\label{spectrum2}
		\end{figure}

In the case of the Hamiltonian (\ref{MagneticHam}), we generally have four Dirac points in the vicinity of $\delta=0$ such that
one expects a four-fold degeneracy of the zero-energy mode for low magnetic fields, in agreement with Eq. (\ref{winding_lowB}).
Because of the parameter $\beta\propto \sqrt{B}$ [see
Eq. (\ref{def:beta})], the low-field limit corresponds to small values of $\beta$, such as in the case of the value $0.06$ chosen to
calculate the spectrum in Figs. \ref{spectrum} and \ref{semiclassical}. Indeed the fourfold degeneracy of the zero mode is lifted
only when approaching the merging transitions from 4 to 2, where $\beta$ diverges as a consequence of the decreasing reciprocal-space
distance between some of the Dirac points. However, this degeneracy can also be lifted exactly at $\delta=0$ by
increasing the value of $\beta$, where Fig. \ref{map} and
Eq. (\ref{winding_highB}) indicate that the degeneracy of the zero-energy mode is 2 above
a certain magnetic field. The quantum-mechanical LL spectra, for larger values of $\beta$, are depicted in Fig. \ref{spectrum2}(a)
for $\beta=0.2$ and for $0.45$ in Fig. \ref{spectrum2}(b), as a function of $\delta$.
In both cases, one notices that the fourfold degeneracy is indeed lifted for all values of $\delta$, in agreement with the
expectation from magnetic blurring. The effect is also apparent in Fig. \ref{spectrum2}(c), where we have plotted the
$\delta=0$ LL spectrum in units of the saddle-point energy $E_S$ as a function of $\beta$. Indeed two branches of
the small-$\beta$ zero-energy mode float away -- due to particle-hole symmetry, one increases in energy while the other one
decreases -- while two other branches are topologically protected and remain at zero energy.

Notice that the magnetic blurring in reciprocal space may also be understood as a blurring in energy. Indeed, the
commutation relations (\ref{eq:CommRel}) induce, via the maps (\ref{eq:map1}), commutation relations for the pseudospin
components that read, to lowest order in $l_B^2$,
\beq\label{eq:CommRelH}
\left[h_x,h_y\right]\simeq \frac{i}{l_B^2}\left(\frac{\partial h_x}{\partial \Pi_y}\frac{\partial h_y}{\partial \Pi_x}
- \frac{\partial h_x}{\partial \Pi_x}\frac{\partial h_y}{\partial \Pi_y}\right).
\eeq
In the vicinity of a Dirac point $j$ with linear band dispersion and a characteristic (possibly anisotropic) Fermi velocity
$(v_{x,j},v_{y,j})$, the commutation relations (\ref{eq:CommRelH}) thus induce a Heisenberg uncertainty relation
$\Delta h_x\Delta h_y\sim v_{x,j}v_{y,j}/l_B^2=v_j^2/l_B^2$ that is precisely on the order of the energy gap between the zero-energy
level and the first excited one. In this picture, the topological winding numbers and thus the level degeneracies associated
with individual contours around the Dirac points are well-defined as long as the energy uncertainty $\sqrt{\Delta h_x\Delta h_y}\sim
v_j/l_B$ is smaller than the saddle point $E_S$. This argument agrees with the expectation that the zero-mode degeneracy
is lifted once
\beq\label{eq:crit0}
E_S\lesssim \sqrt{2}c\sqrt{eB}~~\Leftrightarrow~~\beta\gtrsim \frac{1}{4\sqrt{2}},
\eeq

	\begin{figure}[h!]
	\includegraphics[width=.49\columnwidth]{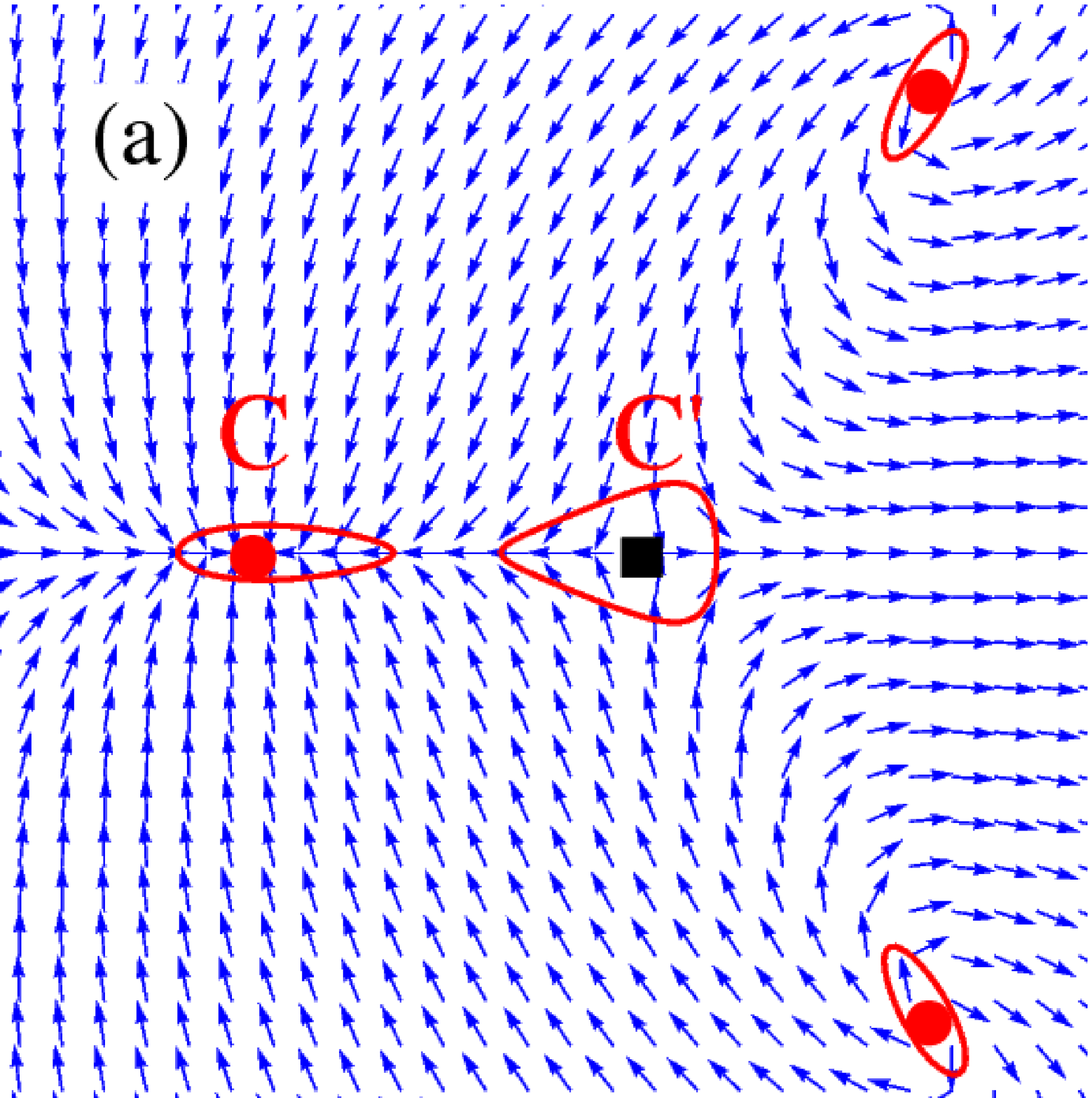}
	\includegraphics[width=.49\columnwidth]{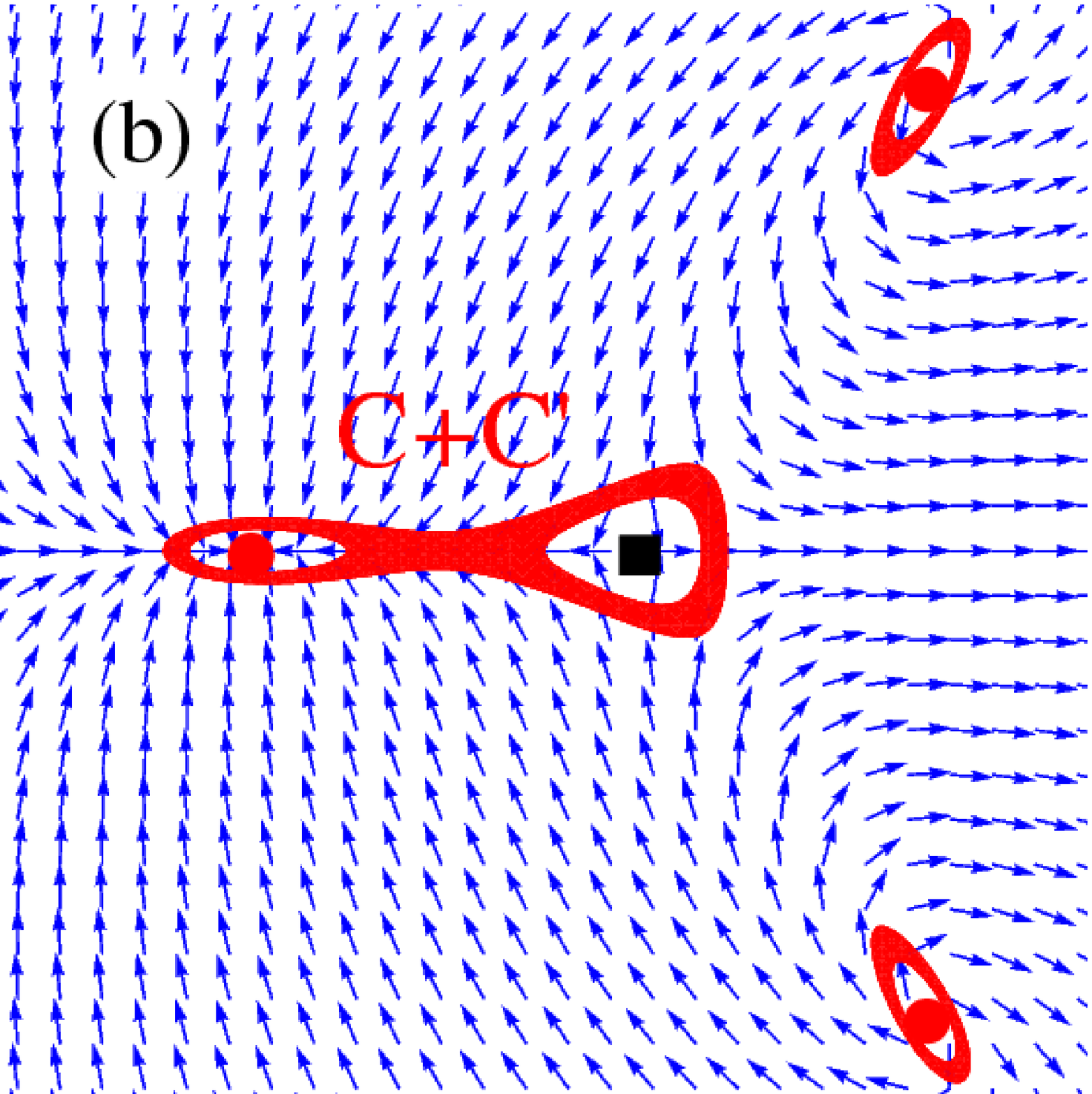}
	\caption{(Color online) Magnetic blurring for contours at higher energy. (a) At low magnetic fields, the semiclassical quantization 
rule holds for each contour encircling a Dirac point. 
(b) When the field increases, the energy contours become blurred and tunneling to trajectories enclosing two singularities becomes possible
in the vicinity of the saddle points. 
The effective number experienced by the electron is $w_{(\mathcal{C}+\mathcal{C'})}=w_{\mathcal{C}}+w_{\mathcal{C'}}=0 $ for the upper bound
of the contour, whereas it is $|w_{\mathcal{C}}|+|w_{\mathcal{C}}|=2$ for the lower bound. }
	\label{BlurringSP}
	\end{figure}

In addition to the zero-energy modes, magnetic blurring also plays a role in the degeneracy lifting of higher-energy LLs in the
vicinity of the saddle points, where the semiclassical approximation does not accurately describe the LL spectrum [see
Fig. \ref{semiclassical}]. Indeed, the degeneracy lifting in the semiclassical approximation is abrupt because of the abrupt
change in the winding number: for energies just below the saddle points, one has disconnected energy contours $\mathcal{C}$ and
$\mathcal{C}^{\prime}$ that
become connected by a contour $\mathcal{C}+\mathcal{C}^{\prime}$ for infinitesimal energies above the saddle points. However, this abrupt transition
is blurred because not only the above-mentioned smallest contours, which are responsible for zero-energy modes, need to enclose
a minimal surface of $\sim 1/l_B^2$, but also two contours corresponding to successive energy levels (see Fig. \ref{BlurringSP}).
The resulting uncertainty about whether a contour in the red region in Fig. \ref{BlurringSP} is connected or disconnected yields
an uncertainty in the winding number, such that the variation of the LLs in the vicinity of saddle points is smoother than that
expected from the semiclassical analysis.

		\section{Detailed analysis of the spectrum}\label{sec:discussion}

The semiclassical and topological theories presented in Secs. \ref{sec:semicl} and \ref{sec:topology}, respectively, allow us to
discuss in detail the different properties of the LL spectrum in Fig. \ref{spectrum}.
From the semiclassical quantization (\ref{eq:ons}) with appropriate values of the winding number $w_\mathcal{C}$,
 we obtain the semiclassical spectrum in the different energy regions separated by the saddle-point energies. Unless stated explicitly, we discuss only
the orbital degeneracy for a single valley ($K$ or $K'$) and a single spin -- the full degeneracy is then given by the orbital
degeneracy times the fourfold spin-valley degeneracy.

			\begin{figure}[h!]
			\includegraphics[width=0.7\columnwidth]{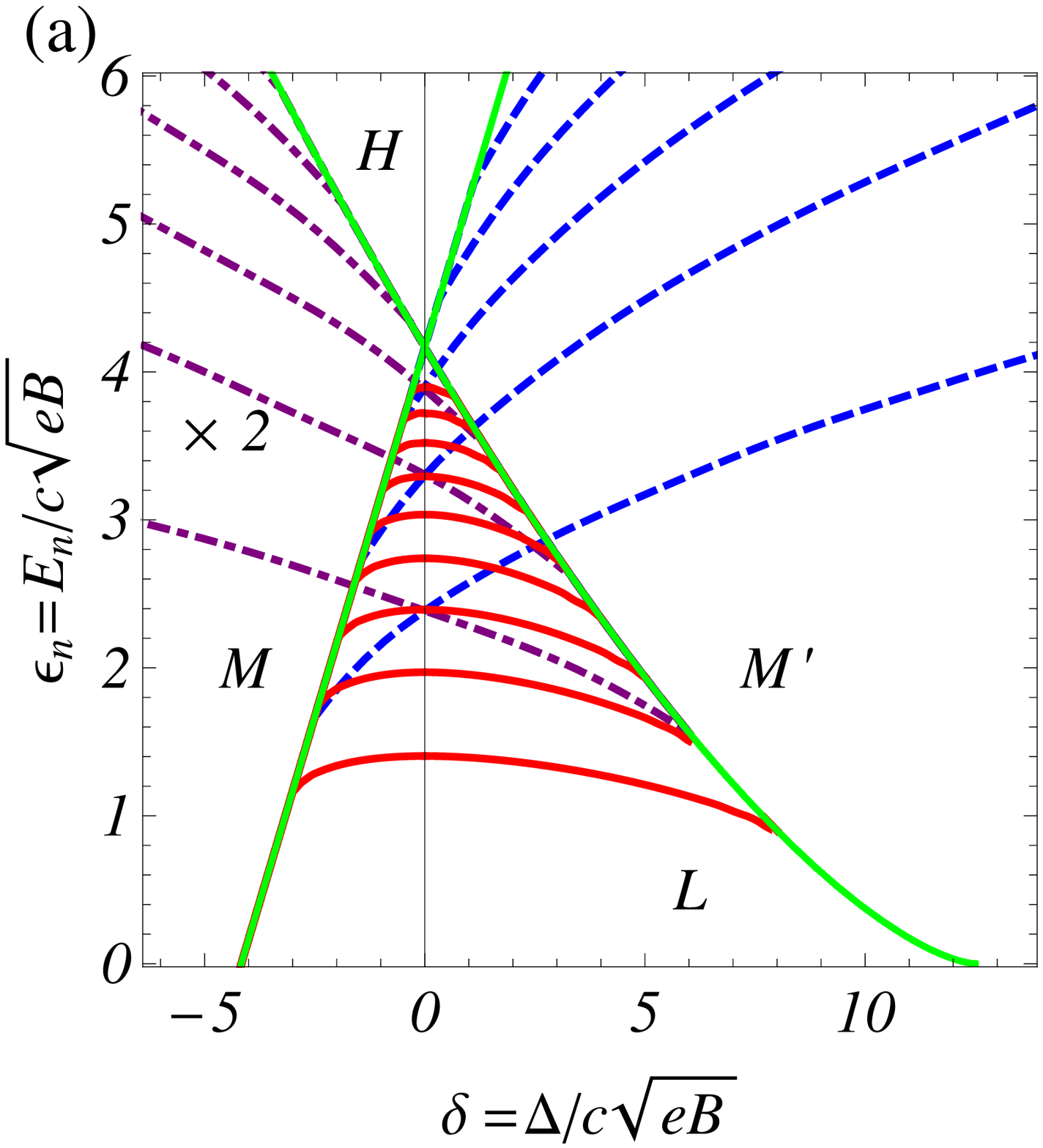}
			\includegraphics[width=0.7\columnwidth]{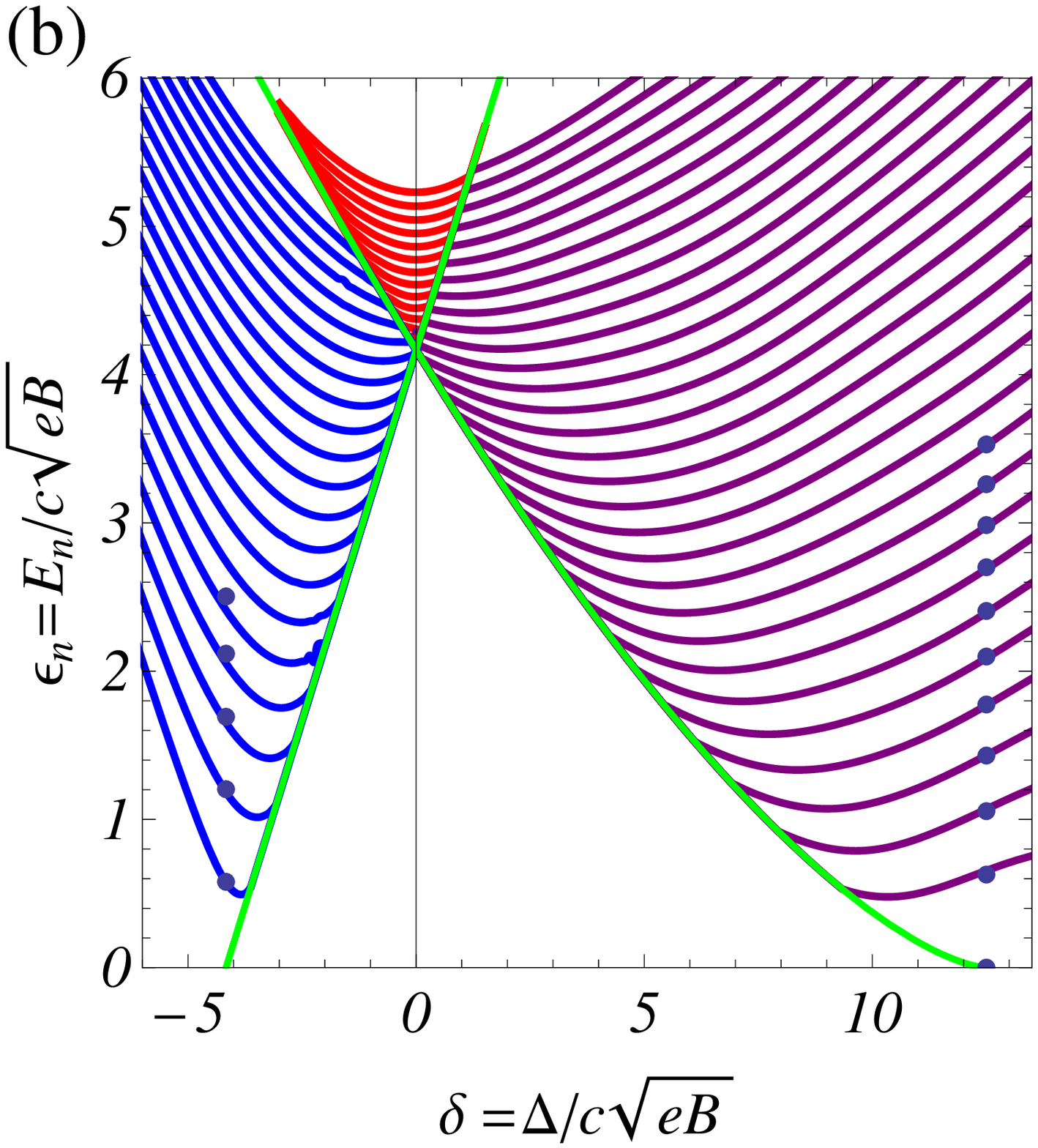}
			\caption{(Color online) Different sectors of the LL spectrum, for $\beta=0.06$.
			(a)  Landau levels in the low-energy sector (L) obtained from the semiclassical approximation ${\cal A}=2 \pi eB  n$.
The red levels correspond to the quantization of the $D$ cone, the blue dashed levels to the quantization of the $A$ cone.
The last set (purple dashed-dotted) of levels is twofold degenerate since it corresponds to the quantization of the two cones labelled $B$ and $C$.
(b) Landau levels in the high-energy sector (above the saddle points). The blue levels in the $M$ region are obtained from the
semiclassical approximation ${\cal A}=2 \pi eB  (n+1/2)$ corresponding to the absence of winding number.  The purple levels in the $M'$ zone
are obtained from the condition ${\cal A}=2 \pi eB  n$ of the pocket issued form the $B, C, D$ cones. The  blue  dots indicate the LL spectrum at the merging (left) and the triple-merging
transition (right), in which case the LLs scale as $(n+1/2)^{2/3}$ and $n^{3/4}$, respectively.  In the high energy region $H$, the four
Dirac pockets have merged into a single pocket with a total winding number $2$, so that the red levels are obtained from the condition
${\cal A}=2 \pi eB  (n+1/2)$.}
			\label{zoneslandau}
			\end{figure}

			\subsection{Undistorted Case $\delta = 0$}\label{sec:spect0}
We begin our reconstruction of the spectrum by plotting the LL at $\delta=0$ ($\Delta = 0$), that is without any distortion, see
vertical lines in Figs. \ref{zoneslandau}(a) and (b).

As long as the energy satisfies $\epsilon < \epsilon_S = \epsilon_{S'}$, the LLs lie in the central region of the low-energy sector (L), see
Fig. \ref{zoneslandau}(a).
In the absence of a magnetic field, the low-energy spectrum is that of Fig. \ref{band0}(a) and consists of four Dirac cones, $A$ to $D$.
The four Dirac cones give rise to four disconnected Fermi pockets the area of which is
\beq
\mathcal{A}_0 = \pi \frac{\epsilon^2}{c^2} \qquad \mathcal{A}_p = \pi \frac{\epsilon^2}{3 c^2},
\eeq
where $\mathcal{A}_0$ is the area of the central cone ($D$) and $\mathcal{A}_p$ that of the other peripheral ones.
In the vicinity of each isolated Dirac cone, one has a topological charge of $w_\mathcal{C}=\pm 1$, as in the case of monolayer graphene, such
that Onsager's quantization rule (\ref{eq:ons}) yields
\beqn
2 \pi \left(n+\frac{1}{2}-\frac{|w_\mathcal{C}|}{2}\right)eB = 2 \pi n e B = \mathcal{A}(E) \\
\nn
\Rightarrow E_0(n) = c\sqrt{2neB} \qquad E_p(n) = c \sqrt{6neB}.
\label{Onsager1}
\eeqn
At zero energy  and for moderate magnetic fields ($\beta=b\sqrt{eB}/c\ll 1$, as in Fig. \ref{spectrum}),
one thus obtains a fourfold orbital degeneracy of the zero-energy level because each of the four Dirac cones yields an $n=0$ LL.
Therefore the total degeneracy of the zero-energy level, taking into account again the additional fourfold spin-valley degeneracy, is 16.
From the topological point of view, this is
related to the presence of four well-separated Dirac points that are not yet blurred by the relatively moderate magnetic field,
$\beta\ll 1$, in such a manner that the four related winding numbers are decoupled and give rise to four zero-energy LLs.

For higher (relativistic) LL, we find three times more LLs for the central cone because $\mathcal{A}_0 = 3\mathcal{A}_p$.
This explains the fourfold degeneracy displayed one time out of three in Fig. \ref{zoneslandau}(a).   Indeed, every third LL is not only
associated with the central Dirac point but also with the three peripheral ones; hence its (accidental) fourfold orbital degeneracy, whereas the other
LLs are non-degenerate because they are associated with the central one only. 
As we have previously discussed [see Eq. (\ref{eq:DispD0})], the averaged Fermi velocity at the points $A$, $B$, and $C$ is $\sqrt{3}$ times larger than that, $c$, around the $D$ point, $v_A = v_{B/C} = \sqrt{3} v_D$.
The three times denser LL spectrum associated with the central cone is therefore a consequence of this relation between the
Fermi velocities and of the approximate LL dispersion
\beq\label{eq:LLdisp}
E_{\pm, n} = \pm\frac{v_j}{l_B}\sqrt{2n}
\eeq
around the Dirac points, with their typical $\sqrt{n}$ scaling, for $j=A$, $B$, $C$, or $D$.

Crossing the transition lines at $\epsilon = \epsilon_S = \epsilon_{S'}$, the spectrum undergoes a transition to the high-energy
sector (H), see Fig. \ref{zoneslandau}(b).
At zero magnetic field, the three peripheral
Fermi pockets merge with the central one.  This change in the topology of the Fermi surface has consequences
for the degeneracy, in the sense that Onsager's quantization rule indicates that there is only one set of LLs associated with the simply connected
Fermi surface. 
There is thus no more orbital LL degeneracy and, besides the trigonal deformation of the Fermi surface, the band structure is approximately parabolic
such that the LLs scale as $n$, as in the usual description of Bernal-stacked bilayer graphene ($\sqrt{n(n-1)} \sim  n -1/2$).

			\subsection{Slightly distorted case $0<|\delta| \ll 1$}

 For small non-zero values of the parameter $\delta$, the saddle points occur at two different energies,
$\epsilon_S\neq\epsilon_{S'}$, and one therefore needs to distinguish three different energy sectors, that is L, M, and H for
$\delta<0$ and L, M$^{\prime}$, and H for $\delta>0$ [see Fig. \ref{zonesberry}].
Below the energies $\epsilon_S$ and $\epsilon_{S'}$, the band structure is that of Fig. \ref{band}(b) or (e), comprising four Dirac cones albeit
with no trigonal symmetry due to finite distortion of the bilayer.  For small values of $\beta$, the picture obtained
in the discussion of the $\delta=0$ case remains essentially unaltered at zero energy. The presence of four distinguishable Dirac cones
yields a fourfold zero-energy level that is insensitive to the slight geometric deformation of the perfectly trigonally-warped case.	
As a consequence, the zero modes remain untouched over a wide range of $\delta$ distortion around $0$.			
		
On the other hand, the higher LL are not topologically protected and the breaking of the trigonal symmetry induces an immediate lift of 
orbital degeneracy, as pictured in Fig. \ref{zoneslandau}.
  Our choice of a real valued parameter $\Delta$ implies that the Dirac cones $B$ and $C$ are related by mirror symmetry, such
that $v_B=v_C$. The corresponding LLs [thick blue lines in Fig. \ref{zoneslandau}(a)] are therefore twofold degenerate and experience
the strongest decrease in energy with increasing $\delta>0$ because their average Fermi velocity is decreased [see Eq. (\ref{eq:FermiV})].
The other two sublevels have a single orbital degeneracy, corresponding to the Dirac cones $A$ and $D$.
  As one may see from Eq. (\ref{eq:FermiV}), the Fermi velocity of the central cone $D$ decreases (quadratically in $\delta$), whereas
that of the cone $A$ increases linearly in $\delta$. As a consequence, the energy of the LLs associated with $D$ [red lines in Fig.
\ref{zoneslandau}(a)] is decreased both for positive and negative values of $\delta$, whereas the $A$-cone LLs increase linearly in
energy with increasing $\delta$.

Above both saddle points, i.e. in the sector H, varying $\delta$  always yields a decrease in the size of the unique Fermi surface
so that the non-degenerate LL are enhanced in energy, as one may see in Fig. \ref{zoneslandau}(b).

			\subsection{Merging transition $\delta \ll -1$}
\label{sec:merging.transition}

In Sec. \ref{sec:topologyB}, we have shown that the fate of the zero-energy level is determined by the parameter $\beta$ -- upon
increase of $\beta$, one obtains a magnetic-field-induced Lifshitz transition from a fourfould degenerate to a twofold degenerate
level. Whereas this picture is roughly the same for small values of $|\delta|$, it needs to be modified when approaching the zero-field
merging transition, that is when the saddle point energy $\epsilon_S$ vanishes. As one may see from Fig. \ref{spectrum}, one notices a significant
departure from the semi-classical approximation. The $A$ and $D$ Fermi pockets merge indeed into a single one and the
corresponding orbital degeneracy of the LLs is changed. 
Indeed, because of the decrease in energy of the saddle point $E_S$, the latter is only higher in energy than the typical scale
$\sqrt{2}c\sqrt{eB}$ for the separation between the lowest LLs if
\beq\label{eq:crit}
1\lesssim \frac{1}{4\sqrt{2}\beta}+\frac{\delta}{\sqrt{2}}.
\eeq
This is a generalization of the criterion (\ref{eq:crit0}) for the undistorted case $\delta=0$.
Based on the criterion (\ref{eq:crit}), one therefore expects the zero-mode degeneracy to be partially lifted at $\delta\simeq \sqrt{2}-1/4\beta$ [that
is $\delta\sim -3$   for our above choice $\beta=0.06$], in good agreement with the numerical results depicted in Fig. \ref{spectrum}. 
Directly at the merging transition, that is for $\delta = -c/4b\sqrt{eB}$, one obtains a LL spectrum with
levels that scale as $\epsilon_n= 2 A \beta^{1/2} (n+1/2)^{2/3}$ with $A= \pi [3 / \Gamma(1/4)]^{2/3} \simeq 1.173$,   in agreement with the merging transition of Dirac cones with opposite Berry phases.\cite{Gilles1,Gilles2}
Upon a further decrease of $\delta$, the merging of the cones $A$ and $D$ is associated with a gap opening [see Fig. \ref{band}(b)]
such that the corresponding LL spectrum is shifted to higher energies, as may be seen on the left-hand side in Fig. \ref{zoneslandau}(b).
Apart from the shift to higher energies, these LLs corresponding to the merged points scale linearly in the LL index $n$, as
one expects for parabolic bands ($\mathcal{A}(E) \propto E \propto n$). Because of the distance in energy from the saddle points, the
semiclassical approximation agrees well with the numeric spectrum, as can be checked in Fig. \ref{semiclassical}.

			\subsection{Triple-merging transition $\delta \gg 1$}
\label{sec:triple.merging}

In the opposite limit, for $\delta >0$, the $A$ cone remains apart and its Fermi velocity $v_A$ is increased. The energy of the LLs
is therefore enhanced and well described within the semi-classical approximation, as may be seen in Fig. \ref{semiclassical}. The $A$ cone
is unaffected by the transition line indicating the saddle point $\epsilon_{S'}$ because it
is not involved in the triple-merging process, as opposed to $B$, $C$ and $D$, that form a boomerang-shaped Fermi surface.
The latter become coupled through the Lifshitz transition at $E_{S'}$, as can be observed from the departure from the semiclassical 
approximation in Fig. \ref{semiclassical}.
Before this transition,  all three Fermi pockets increase in size, with a higher rate for $B$ and $C$ than for $D$, such that the
twofold degenerate LLs corresponding to the points $B$ and $C$ decrease faster in energy than those of the central cone $D$.

There are only two zero-energy LLs since the total $+1$ topological charge of the boomerang pocket gives rise  to a unique topologically protected mode.
Equivalently, the magnetic field has reached such a value that the total winding number $w_p$ is the only relevant quantity.

Precisely at the triple merging point ($E_{S'}=0$), the LL scale as $n^{3/4}$ [blue dots in Fig. \ref{spectrum}(b)], as far as our numerical accuracy
is concerned.
After the triple merging transition, the LL of $D=B=C$ scale with a different power law, almost linear in $n$.
Increasing $\delta$ increases the energies of all the LL because of a decrease of the combined-orbit area.

			\subsection{LL spectrum for an imaginary value of $\delta$}
		\begin{figure}[h!]
				\includegraphics[width=.49\columnwidth]{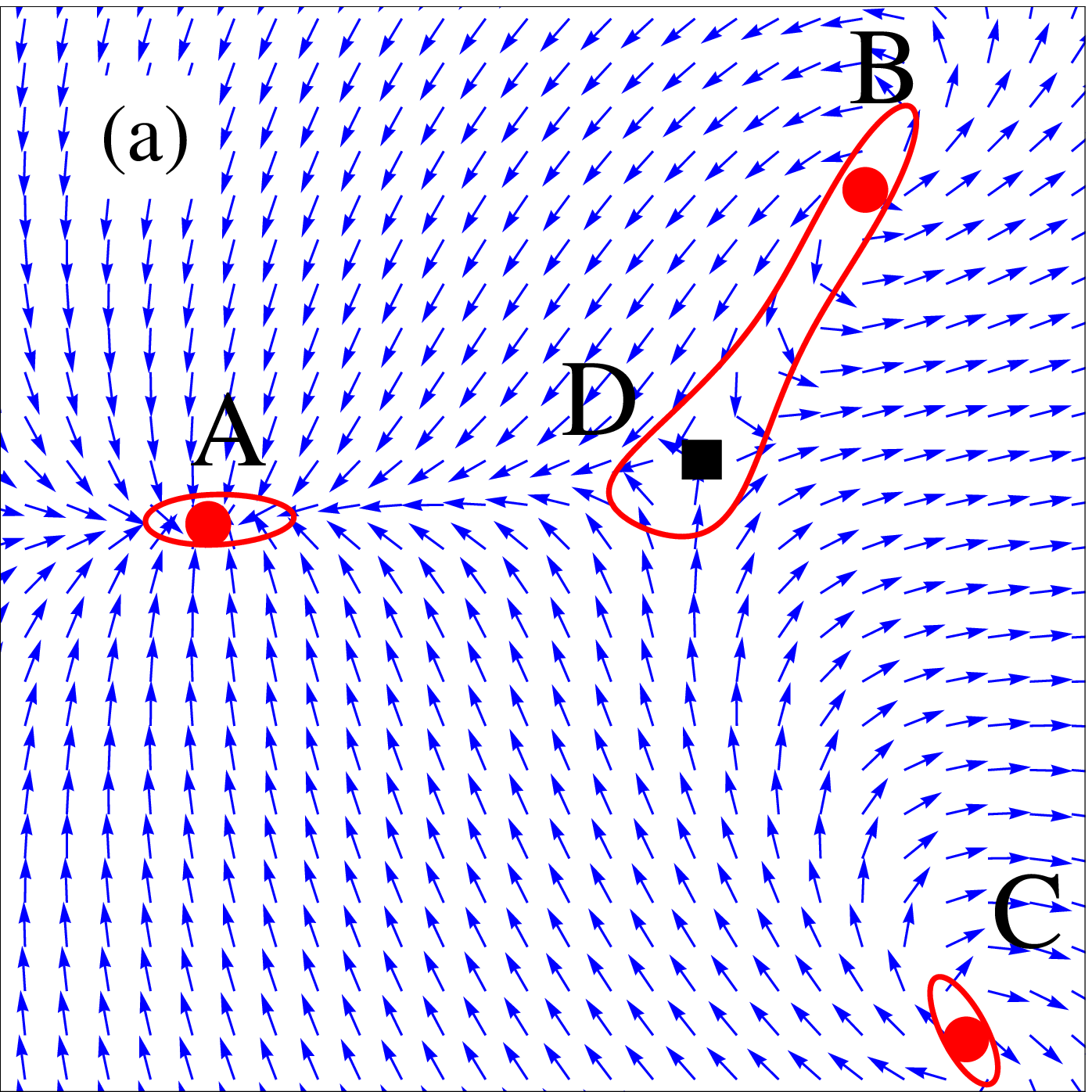}
				\includegraphics[width=.49\columnwidth]{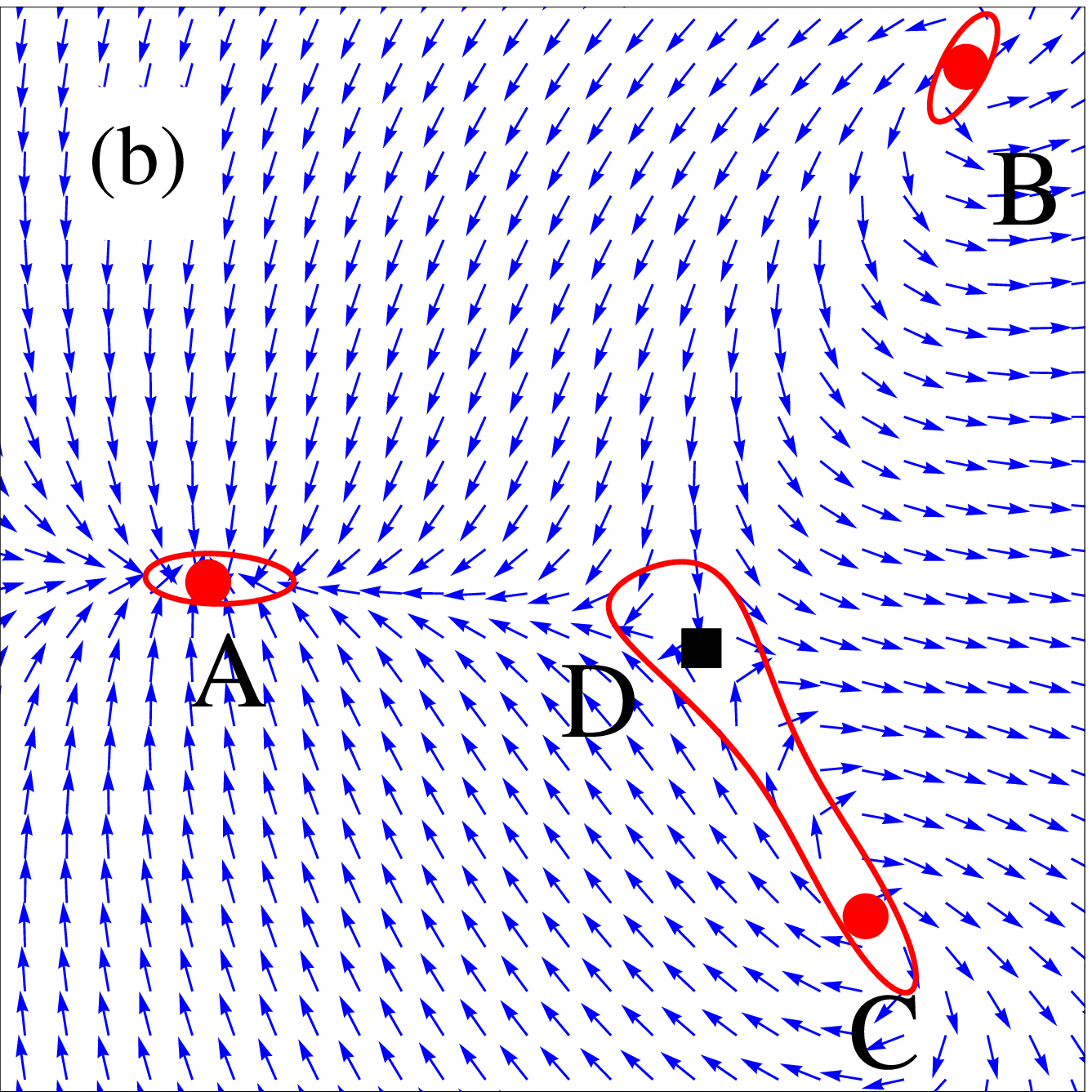}
				\caption{(Color online)  An imaginary value of $\Delta$ ($\theta=\pi/2$)  induces a merging between the $B$ and $D$ cones 
[panel (a) for $\text{Im}(\Delta) >0$) or between the $C$ and $D$ cones [panel (b) for $\text{Im}(\Delta)<0$].}
				\label{deltaim}
				\end{figure}

For the sake of completeness, we present  here a LL spectrum 
for a purely imaginary $\Delta$ in the case of a deformation in the $x$-axis ($\theta=\pi/2$ in Fig. \ref{thetam}, see Sec. \ref{sec:BandStruc} 3).
Then, the two directions of deformation [$\pm \text{Im}(\Delta)$] are equivalent, 
as pictured in Figs.   \ref{thetamp}(b) 
and \ref{deltaim}, so that the saddle point energy and the LL spectrum is now symmetric in $\text{Im}(\Delta)$, as seen in Fig. \ref{thetam}. 
When $\text{Im}(\Delta) >0$, the cones $B$ and $D$ merge leaving the cones $A$ and $C$ isolated, whereas for 
$\text{Im}(\Delta)<0$, the role of the $B$ and $C$ points is interchanged.  In a magnetic field, one   distinguishes the LL sequence 
from the four cones below the saddle point energy, as well as the Lifshitz transition near the saddle points. 
Notice that the accidental degeneracy of the $B$ and $C$ cones, which we have encountered for real values of $\Delta$, 
is now lifted and that the LLs associated with the $A$ cone are
symmetric in $\text{Im}(\Delta)$ (Fig. \ref{thetam}). Indeed, the $A$ cone remains isolated and does not take part in the merging
transition for any value of $\text{Im}(\Delta)$. Its LL spectrum therefore remains relativistic with the typical $\sqrt{n}$ scaling. In the 
merging sector, for energies in between the two saddle points, the $C$ cone provides an additional set of relativistic LLs for $\text{Im}(\Delta)>0$,
whereas this set is provided by the $B$ cone for $\text{Im}(\Delta)<0$. As in the case of the merging transition discussed in 
Sec. \ref{sec:merging.transition}, beyond $|\Delta_m|=(c^2/4b)(6 \sqrt{3} - 9)^{1/2}\simeq 1.18 c^2/4b$
($|\delta_m|\simeq 4.92$, for $\beta=0.06$ as shown in Fig. \ref{deltaim}) the merged cones [$B$ and $D$ for $\text{Im}(\Delta)>
1.18 c^2/4b$ or $C$ and $D$ for 
$\text{Im}(\Delta)<-1.18 c^2/4b$] are accompanied by the opening of a local gap the $\delta$-dependence of which is indicated by the thick blue line
in Fig. \ref{thetam}. Consequently the associated LLs are non-relativistic with a linear-$n$ scaling because of the annihilation of the winding 
numbers of the merged cones.

				\begin{figure}[h!]
				\includegraphics[width=1.0\columnwidth]{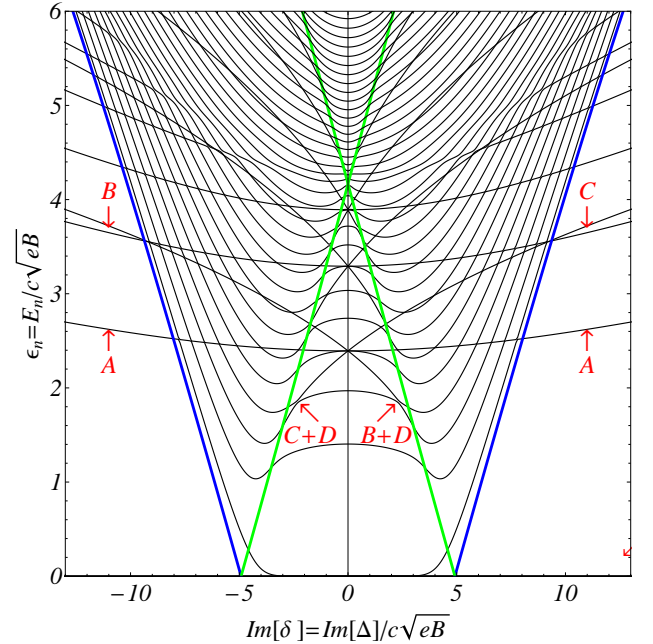}
				\caption{(Color online) Landau level spectrum for $\theta=\pi/2$ ($\beta=0.06$). 
In this case, the spectrum is symmetric in the 
displacement $\text{Im}(\delta)=\text{Im}(\Delta)/c\sqrt{eB}$. The saddle point energies are indicated by the green 					 
curves and the energy of the local gap beyond merging is indicated by the blue curves. }
				\label{thetam}
				\end{figure}

\subsection{Consequences for magneto-transport measurements}

We finish this section with a brief discussion of the consequences of the above picture for magneto-transport measurements, namely in the 
context of Hall quantization. Such experiments have been performed both in bilayer graphene in the low-energy limit,\cite{Novoselov11} as 
well as in samples with a twist between the two layers.\cite{exp_twisted1,exp_twisted2} Remember that the model (\ref{fullHam}) investigated above 
also accounts for the case of twisted bilayer graphene, with moderate twist angles, if one sets $c=0$ or in the limit $c^2/b\ll \Delta$. In this case, 
$\Delta$ is a function of the twist angle. 

In all measurements, an eightfold degeneracy of the zero-energy level, with no additional quantum-Hall plateaus in the range $-4<\nu<4$, has been 
observed. This indicates, in addition to the usual fourfold spin-valley degeneracy, a twofold degeneracy of the zero-energy level. In the case of 
twisted bilayer graphene,\cite{exp_twisted1,exp_twisted2} this is an indication for the twofold topological degeneracy associated with two Dirac points
characterized by a unit winding number with the same sign.\cite{degail11} For a bilayer sample with no twist,\cite{Novoselov11} 
the observed eightfold degeneracy indicates a prominent non-zero value of $\Delta$ since one would expect, based on the above arguments for $\Delta\sim 0$,
a 16-fold degeneracy of the zero-energy level, i.e. no quantum-Hall plateaus in between $-8<\nu<8$. It has been argued that the rather large value 
of $\Delta$ cannot be explained by strain (or a displacement of the two layers) alone and that interaction effects are likely to play an important 
role,\cite{Novoselov11} in which case $\Delta$ plays the role of a (nematic) order parameter.\cite{Vafek}

In higher LLs in trigonally-warped bilayer graphene, 
the degeneracy depends both on the value of the saddle point energy $\sim |\Delta|$ as compared to the magnetic energy scale 
$c\sqrt{eB}$, as well as on the phase $\theta$ of $\Delta=|\Delta|\exp(i\theta)$. In the low-energy sector (L), we have shown that for a real value of
$\Delta$ ($\theta=0$ or $\pi$, modulo $2\pi/3$), the Dirac cones at $B$ and $C$ are related by mirror symmetry and their LLs are thus $(2\times 4)$-fold 
degenerate in the low-energy (L) and merging (M) sectors.
One would therefore expect a jump of $\Delta \nu=8$ in the Hall conductance whenever the Fermi level crosses such a level, whereas the 
LLs associated with the points $A$ and $D$ are only spin-valley degenerate, associated with a jump $\Delta \nu=4$. Notice that the mirror symmetry
is immediately broken in the case of a non-zero imaginary part of $\Delta$, i.e. when $\theta\neq 0$ or $\pi$ (modulo $2\pi/3$), such that all LLs are then
fourfold spin-valley degenerate only. This fourfold degeneracy is also the generic case in the other energy sectors ($M'$ and $H$). Experimentally, 
quantum-Hall features have been observed at $\nu=\pm 4,\pm 8, \pm 12, ...$,\cite{Novoselov11} such that the LLs are only spin-valley degenerate. Whereas
this sequence is identical to that of bilayer graphene without trigonal warping, the $\sqrt{B}$-scaling of the gap between the zero-energy level and the
first excited one indicates that the low-energy sector is nevertheless governed by Dirac cones with a linear dispersion relation, as one would expect 
in the sectors $M$ and $M'$.

The situation is different in twisted bilayer graphene, where one expects eightfold-degenerate LLs below the saddle point at $E\ll |\Delta|$, whereas 
they are fourfold-degenerate at $E\gtrsim |\Delta|$. Since the value $|\Delta|$ can be tuned to great extent by the twist angle, one may expect to see
this crossover more easily than in trigonally-warped bilayer graphene. From an experimental point of view, Lee \textit{et al.}\cite{exp_twisted1}
investigated an epitaxially grown sample on SiC, with typical twist angles of $2.2^{\circ}$. In this sample an eightfold degeneracy of the zero-energy
level has been observed, whereas higher LLs are fourfold spin-valley degenerate, that is a filling-factor sequence of $\nu=\pm 4,\pm 8, \pm 12, ...$.
Sanchez-Yamagishi \textit{et al.} have investigated a twisted bilayer sample fabricated by PMMA-transfer technique of two monolayer samples on 
hexa-boron-nitride.\cite{exp_twisted2} In this case, the observed sequence of quantum-Hall plateaus is $\nu=\pm 4,\pm 12, \pm 20, ...$, that is eightfold-degenerate
Landau levels also at higher energy. This indicates a large value of $|\Delta|$ and of relatively large twist angles.

	\section{Conclusions}\label{sec:concl}
In conclusion, we have investigated a continuum model that accounts for the presence of two and more Dirac points in the dispersion
relation. The model describes the low-energy physical properties of bilayer graphene with a stacking default, either a translational
displacement of one graphene layer with respect to the other one, as compared to the perfectly AB-stacked case, or
strain.\cite{Son,Falko11} Furthermore, this model also accounts for a rotational stacking default (twist) if one neglects the linear
term $\mathcal{H}_c$ in the low-energy Hamiltonian (\ref{fullHam}).\cite{degail11} Whereas the number of Dirac points is determined
by the interplay between the different microscopic parameters, the total winding number of $+2$ topologically guarantees the presence
of at least two Dirac points (with winding number $+1$)
or a single parabolic band-contact point (with winding number $+2$).

In the presence of a magnetic field and Landau quantization, the winding number yields a doubly degenerate zero-energy
level that is topologically protected. We have studied the LL spectrum in the framework of a semiclassical treatment and find
that it describes accurately the numerically obtained one in a large parameter range. The semiclassical theory allows for
a detailed understanding of the LL spectrum in the sense that one may associate certain levels with particular Dirac points and
determine the degeneracy of the levels. Furthermore, the degeneracy lifting is understood in terms of connections between Fermi
pockets.

However, the semiclassical approximation,
which is based on a quantization of reciprocal-space orbits and the topological charge (i.e. the winding number),
breaks down in the vicinity of
saddle points in the (zero-field) dispersion relation as well as at zero energy in the high-magnetic-field limit. The physical
origin if this breakdown is the definition of the topological charges in terms of closed
reciprocal-space orbits, which change abruptly
at the saddle points when two or more Fermi pockets become connected. Indeed, the definition of topological charges needs to be
revisited in the presence of a magnetic field that quantizes reciprocal space into patches of size $\sim 1/l_B^2\propto B$ because the
components of the kinetic-momentum operator no longer commute. This effect blurs the reciprocal-space orbits and smoothens
the abrupt change in the winding number at the saddle points.

Another effect of this magnetic blurring concerns the zero-energy states. Because reciprocal-space orbits need to enclose minimal
surfaces of $\sim 1/l_B^2$, neighboring Dirac points at zero energy are no longer resolved individually in the high-field limit. This
effect is at the origin of a magnetic-field-induced Lifshitz transition, where the degeneracy of the zero-energy level, which consists of
four ($w_t=4$) $n=0$ LLs (associated with the total sum $w_t$ of Dirac points), is partially lifted when increasing the magnetic field.
Eventually, the degeneracy of the zero-energy level is then given by the total topological charge, that is $w_p=|\sum_i w_i|$,
in terms of the zero-field winding number $w_i=\pm 1$ of a single Dirac point.

	\section*{Acknowledgements}
We acknowledge fruitful discussions with Antonio H. Castro Neto, Jean-No\"el Fuchs, Kostya Novoselov, and Fr\'ed\'eric Pi\'echon.
This work was supported by the ANR project NANOSIM GRAPHENE under Grant No.
ANR-09-NANO-016 and by the Ecole Doctorale de Physique de la R\'egion Parisienne (ED 107).

	\bibliography{biblio}
	
\end{document}